\documentclass[preprint,aps,showpacs,preprintnumbers,amsmath,amssymb,nofootinbib,epsf,epsfig]
{revtex4}
\usepackage{graphicx}

\usepackage{epsfig}
\begin{document}

\begin{flushright}
\end{flushright}


\newcommand{\be}{\begin{equation}}
\newcommand{\ee}{\end{equation}}
\newcommand{\bea}{\begin{eqnarray}}
\newcommand{\eea}{\end{eqnarray}}
\newcommand{\nn}{\nonumber}
\def\ds{\displaystyle}
\def\s1{\hat s}
\def\para{\parallel}
\newcommand{\mrm}[1]{\mathrm{#1}}
\newcommand{\mc}[1]{\mathcal{#1}}
\def\CP{{\it CP}~}
\def\cp{{\it CP}}
\def\ml{m_\mu}
\title{\large Scalar leptoquarks and  the rare   $B$ meson decays}
\author{Suchismita Sahoo, Rukmani Mohanta  }
\affiliation{
School of Physics, University of Hyderabad, Hyderabad - 500 046, India }

\begin{abstract}
We study some rare decays of $B$ meson involving the quark level transition
$b \to q l^+l^-~(q=d,s)$ in the scalar leptoquark model. We constrain the leptoquark parameter space 
using the recently measured branching ratios of
$B_{s,d} \to \mu^+ \mu^-$ processes. Using such parameters, we obtain the branching ratios, direct CP violation 
parameters and isospin
asymmetries in $B \to K \mu^+ \mu^-$ and $B \to \pi \mu^+ \mu^-$ processes.
We also obtain the branching ratios for some lepton flavour violating decays $B \to l_i^+ l_j^-$.
We find that the various anomalies associated with the isospin asymmetries of  $B \to K \mu^+ \mu^-$ process
can be  explained in the
scalar leptoquark model.

\end{abstract}
\pacs{13.20.He, 14.80.Sv}

\maketitle

\section{Introduction}

The rare decays of $B$ mesons involving flavor changing neutral current (FCNC) transitions
$b \to s/d$, provide an excellent testing ground to look for new physics.
In the standard model (SM), these transitions occur  at one-loop level
and hence, they are very sensitive to any new physics contributions.
Although, so far we have not seen any clear indication of new physics in  the $b$ sector,
but there appears to be some kind of tension with the SM predictions in some  $b \to s$ penguin induced transitions.
It should be noted that the recent measurement by LHCb collaboration \cite{lhcb} shows several significant deviations
on angular observables in the rare decay $B \to K^{* 0} \mu^+ \mu^-$ from their corresponding SM expectations.
In particular, the most
significant discrepancy of 3.7$\sigma$, arises in the variable $P_5'$  \cite{matias} (the analogue of $S_5$ in \cite{buras})
provides high sensitivity to new physics (NP) effects in $b \to s \gamma,  ~sl^+ l^-$ transitions. Further
results from LHCb experiment  in combination  with the critical assessment of the theoretical
uncertainties  will be necessary to
clarify whether the observed deviations are a real sign of
NP or simply the statistical fluctuations \cite{martin,hurth}.

Another  indication of new physics is related to the recent measurement of isospin asymmetry in
$B \to K \mu^+ \mu^-$ process by LHCb experiment \cite{lhcb1}, which gives  a negative
deviation from zero  at the level of $4\sigma$  taking into account the entire
$q^2$-spectrum. The isospin-asymmetry in $B \to  K ll$ is expected to be vanishingly small in the SM and
hence, the measured asymmetry provides another smoking-gun signal for new physics.

More recently another discrepancy occurs in the  measurement of the ratio of branching fractions
of $B \to Kl^+l^- $ decays into dimuons over dielectrons by the LHCb collaboration \cite{lhcb2},
\bea
R_K = \frac{{\rm BR}( B \to K \mu^+ \mu^-)}{{\rm BR}( B \to K e^+ e^-)}\;,
\eea
and the obtained value  in the dilepton invariant mass squared bin $(1 \lesssim q^2 < 6)~ {\rm GeV^2}$ is
\bea
R_K^{LHCb}=0.745_{-0.074}^{+0.090} \pm 0.036\;.
\eea
Combining the  statistical and systematic uncertainties in quadrature,
this observation corresponds to a $2.6 \sigma$  deviation from its SM prediction
$R_K = 1.0003 \pm 0.0001$ \cite{hiller}, where corrections of order $\alpha_s$  and  $(1/m_b)$
are included. In contrast to the anomaly in the rare decay
$B \to  K^* \mu^+ \mu^-$, which is affected by unknown power corrections,
the ratio $R_K$ is theoretically clean and  this
might be a sign of  lepton flavour non-universal physics.

Although it is conceivable that these anomalies mostly associated with $b \to s l^+l^-$ transitions are
due to statistical fluctuations or under-estimated
theory uncertainties, but the possible interplay of new physics could not be ruled out. These LHCb  results 
 have attracted many theoretical attentions in recent times \cite{matias,martin,hurth,np} both in the context of some new physics model 
 or in  model independent way.
In this paper we would like to investigate some of the rare decay modes of $B$ meson involving the
FCNC transitions $b \to (s,d)l^+l^-$, e.g., $ B \to K l^+l^-$,
$B \to \pi l^+ l^-$ and $B \to l_i^+ l_j^-$ using the scalar leptoquark (LQ) model. In particular, we would like to
see whether the leptoquark model can accommodate some of the anomalies discussed above, in particular the ones associated with 
$B \to K ll$ processes. It is well-known that 
leptoquarks are color-triplet bosonic particles that can couple to a quark and a lepton at the same time and can occur
in various extensions of the standard model \cite{lepto}.  They can have spin-1 (vector leptoquarks), which exist in grand unified theories based on
$SU(5)$, $SO(10)$ etc., or spin-0 (scalar leptoquarks).  Scalar leptoquarks can exist at  TeV scale in
extended  technicolor models \cite{lepto1} as well as  in  quark and lepton composite models \cite{lepto2}.
The phenomenology of scalar leptoquarks have been studied extensively in the literature \cite{lepto4,wise,lepto5}.
It is generally assumed that the vector leptoquarks tend to couple directly to neutrinos, and hence it is expected 
that their couplings are tightly constrained from the neutrino mass and mixing data. Therefore, 
in this paper we  consider the model where leptoquarks can couple only to a pair of quarks and leptons and hence
may be inert with respect to  proton decay. Hence, the bounds from proton decay may not be applicable for such cases and
leptoquarks may produce signatures in other low-energy phenomena \cite{wise}.

The paper is organized as follows. In section II we briefly discuss the effective Hamiltonian describing the process $b \to s l^+ l^-$
and the new contributions arising due to the exchange of scalar leptoquarks. The constraints on the
leptoquark parameter space are obtained using the recently measured branching ratios of the decay modes 
$B_{s,d} \to \mu^+ \mu^-$  and $ B \to X_s e^+ e^-$ process in sections III.
The branching ratios and various asymmetries of the rare decay modes $B \to K l^+l^-$
and $B \to \pi l^+l^-$ are discussed in sections IV and V respectively. In Section VI we present
the lepton flavour violating decays $B \to l_i^+ l_j^-$ and Section VII contains the conclusion.
\section{Effective Hamiltonian for $b \to s l^+ l^- $ process}

In the standard model effective Hamiltonian describing the quark level transition $b \to s ll$ is given as \cite{buras1}
\bea
{\cal H}_{eff} &=& - \frac{ 4 G_F}{\sqrt 2} V_{tb} V_{ts}^* \Bigg[\sum_{i=1}^6 C_i(\mu) O_i +C_7 \frac{e}{16 \pi^2} \Big(\bar s \sigma_{\mu \nu}
(m_s P_L + m_b P_R ) b\Big) F^{\mu \nu} \nn\\
&&+C_9^{eff} \frac{\alpha}{4 \pi} (\bar s \gamma^\mu P_L b) \bar l \gamma_\mu l + C_{10} \frac{\alpha}{4 \pi} (\bar s \gamma^\mu P_L b)
\bar l \gamma_\mu \gamma_5 l\Bigg]\;,\label{ham}
\eea
where   $G_F$ is the Fermi constant and $V_{q q'}$ are the Cabibbo-Kobayashi-Maskawa (CKM)
matrix elements, $\alpha$ is the fine-structure constant, $P_{L,R} = (1 \mp \gamma_5)/2$ and $C_i$'s are the Wilson coefficients.
The values of the Wilson coefficients are calculated at the next-to-next-leading order (NLL) by matching the full theory
to the effective theory at the electroweak scale and subsequently solving the renormalization group equation (RGE) to run them down
to the $b$-quark mass scale  \cite{buras}, and the values used in this analysis are listed in Table-1.

\begin{table}[ht]
\begin{center}
\vspace*{0.1 true in}
\begin{tabular}{c c c c c c c c c c  }
\hline
\hline
 $C_1$ & $C_2$ & $C_3$ & $C_4$ & $C_5$ & $C_6$ & $C_7^{eff}$ & $C_8^{eff}$ & $C_9$ & $C_{10}$ \\
\hline
$-0.3001 $ &~ $1.008 $ ~& $-0.0047 $ ~&~ $-0.0827 $ ~&~ $0.0003 $~ &~ $0.0009 $ ~& $-0.2969 $ 
~& $-0.1642$~ &~ $4.2607 $ ~&$-4.2453 $\\
\hline\hline
\end{tabular}
\end{center}
\caption{The SM Wilson coefficients evaluated at the scale $\mu=4.6$ GeV \cite{hou}.}
\end{table}
\subsection{New Physics Contributions due to Scalar Leptoquark exchange}

The effective Hamiltonian  (\ref{ham}) will be modified in the leptoquark model  due
to the additional contributions arising from the exchange of scalar leptoquarks. Here, we will consider the minimal renormalizable scalar
leptoquark models \cite{wise}, containing one single additional representation of $SU(3) \times SU(2) \times U(1)$ and which  
do  not allow proton decay at the tree level. It has also  been shown  that 
this requirement can only be satisfied by two models and in these models, the leptoquarks can
have the representation as  $X=(3,2,7/6)$ and $X=(3,2,1/6)$ under the gauge group $SU(3) \times SU(2) \times U(1)$.
Our objective  here is to consider these scalar leptoquarks which potentially contribute to the $b \to (s,d) \mu^+ \mu^-$ transitions
and constrain the underlying couplings from experimental data on $B_{s,d} \to \mu^+ \mu^-$.
The details of these new contributions are  explicitly discussed in Ref. \cite{rm1}, and here we simply
outline the main points.

The interaction Lagrangian  for the coupling of scalar leptoquark $X= (3,2,7/6)$ to the fermion bilinears is given as
\bea
{\cal L}= -\lambda_u^{ij}~ \bar u_{\alpha R}^i ( V_\alpha e_L^j - Y_\alpha \nu_L^j )
-\lambda_e^{ij}~ \bar e_R^i \left (V_L^\dagger u_{\alpha L}^j + Y_\alpha^\dagger d_{\alpha L}^j \right )+h.c.\;.\label{lepto}
\eea
Using the Fierz transformation,  one can obtain from Eq. (\ref{lepto}),
 the contribution to the interaction Hamiltonian for
the  $b \to s \mu^+ \mu^- $ process as
\bea
{\cal H}_{LQ}&=& \frac{\lambda_\mu^{32} {\lambda_\mu^{22 }}^* }{8 M_Y^2} [ \bar s \gamma^\mu (1-\gamma_5)b]
[\bar \mu \gamma_\mu(1+\gamma_5) \mu ]\equiv \frac{\lambda_\mu^{32} {\lambda_\mu^{22 }}^*}{4 M_Y^2} \Big (O_9 +O_{10} \Big).
\eea
One can thus write the leptoquark effective Hamiltonian (5) analogous to its  SM counterpart (\ref{ham}) as
\bea
{\cal H}_{LQ}=- \frac{  G_F \alpha}{\sqrt 2 \pi} V_{tb} V_{ts}^*(C_9^{NP} O_9 +C_{10}^{NP} O_{10})\;,
\eea
with the new Wilson coefficients
\bea
C_9^{NP} = C_{10}^{NP} = - \frac{ \pi}{2 \sqrt 2 G_F \alpha V_{tb} V_{ts}^* }\frac{\lambda_\mu^{32}{ \lambda_\mu^{22 }}^*}{
M_Y^2}\;.\label{c10np}
\eea
Similarly the interaction Lagrangian for the coupling of
$X=(3,2,1/6)$ leptoquark to the fermion bilinear can be expressed as
 \bea
{\cal L} = - \lambda_d^{ij}~ {\bar d}_{\alpha R}^{~i} (V_\alpha e_L^j-Y_\alpha \nu_L^j) +h.c.\;,
\eea
and after performing the Fierz transformation, the interaction Hamiltonian becomes 
\bea
{\cal H}_{LQ}&=& \frac{\lambda_s^{22} {\lambda_b^{32}}^*}{8 M_V^2}[\bar s \gamma^\mu (1+\gamma_5) b] [ \bar \mu \gamma_\mu (1- \gamma_5) \mu ]
= \frac{\lambda_s^{22} {\lambda_b^{32}}^*}{4 M_V^2}\left ( O_9^{'NP }-O_{10}^{'NP } \right )\;,
\eea
where $O_9'$ and $O_{10}'$ are  the four-fermion current-current operators  obtained from $O_{9,10}$ by making the replacement $P_L \leftrightarrow P_R$.
Thus,  due to the exchange of the leptoquark $X=(3,2,1/6)$, one can obtain  the new  Wilson coefficients $C_9^{'NP }$  and  $C_{10}^{'NP }$ associated 
with the operators $O_9'$ and $O_{10}'$ as  
\bea
C_9^{'NP } = - C_{10}^{'NP } = \frac{ \pi}{2 \sqrt 2 ~G_F \alpha V_{tb}V_{ts}^*} \frac{\lambda_s^{22} {\lambda_b^{32}}^*}{M_V^2}\;.\label{c10np1}
\eea
The analogous new physics contributions for $b \to d\mu^+ \mu^-$ transitions can be obtained from
$b \to s \mu^+ \mu^-$ process by replacing the leptoquark couplings $\lambda^{32} {\lambda^{22}}^*$ by $\lambda^{32} {\lambda^{12}}^*$ and the CKM
elements $V_{tb} V_{ts}^*$ by $V_{tb} V_{td}^*$ in Eqs. (6-10).
After having the idea of  new physics contributions to the process $ b \to (s,d) \mu^+ \mu^-$, we now proceed to constrain the
new physics parameter space using the recent measurement of $B_{s,d} \to \mu^+ \mu^-$.

\section{$B_{s,d} \to \mu^+ \mu^-$ decay process}

The rare leptonic decay processes $B_{s,d} \to \mu^+ \mu^-$, mediated by the FCNC transition
$b \to s,d$ are strongly  suppressed  in the standard model as they occur at one-loop level as well as suffer
from helicity suppression. These decay processes are
very clean and the only nonperturbative quantity involved is
the   $B$ meson decay constant, which can be reliably calculated using the
non-perturbative methods such as QCD sum rules, lattice gauge theory etc.
Therefore, they  are considered as one of the most powerful tools to provide
important constraints on models of new physics.
These processes have been very well studied in the literature and in recent times also 
they have attracted a lot of attention
\cite{fleischer1,fleischer2,buras5,mu6,staub,kosnik, bsll}. Therefore, here we will point out
the main points.
The constraint on the leptoquark couplings from $B_s \to \mu^+ \mu^-$ are recently 
extracted by one of us in Ref. \cite{rm1}.

The most general effective Hamiltonian
describing these processes is given as
\bea
{\cal{H}}_{eff}
= \frac{G_F \alpha }{\sqrt{2} \pi} V_{tb} V_{tq}^*
\Bigg[C_{10}^{eff}O_{10} + C_{10}' O_{10}' \Bigg],\label{hammu}
\eea
where $q=d$ or $s$, $C_{10}^{eff} = C_{10}^{SM} + C_{10}^{NP}$ and $C_{10}'= C_{10}^{'NP}$.
The corresponding branching ratio  is given as
\be
{\rm BR}(B_q \to \mu^+ \mu^-) = \frac{G_F^2}{16 \pi^3} \tau_{B_q} \alpha^2 f_{B_q}^2 M_{B_q} m_{\mu}^2 |V_{tb} V_{tq}^*|^2
\left |C_{10}^{eff}-C_{10}'\right |^2 \sqrt{1- \frac{4 \ml^2}{M_{B_q}^2}}.
\ee
However, as discussed in Ref . \cite{fleischer1}, the average  time-integrated  branching ratios
${\overline{\rm BR}}(B_q \to \mu^+ \mu^-)$ depend on the details of $B_q - \bar{B}_q$ mixing, which
in the SM,  related to the decay widths $\Gamma(B_q \to \mu^+ \mu^-)$
by a very simple relation as ${\overline{\rm BR}}(B_q \to \mu^+ \mu^-)=\Gamma(B_q \to \mu^+ \mu^-)/\Gamma_H^q$,
where $\Gamma_H^q$ is the total width of the heavier mass eigen state.

Including the corrections of ${\cal O}(\alpha_{em})$ and ${\cal O}(\alpha_s^2)$, the updated branching ratios in the
standard model are calculated in \cite{bsll} as
\bea
{\overline{\rm BR}}(B_s \to \mu^+ \mu^-)|_{\rm SM}&=&\left (3.65 \pm 0.23 \right ) \times 10^{-9},\nn\\
{\overline{\rm BR}}(B_d \to \mu^+ \mu^-)|_{\rm SM}
&=&\left (1.06 \pm 0.09  \right ) \times 10^{-10}\;.\label{brmu}
\eea
These processes are recently measured by the CMS \cite{cms} and LHCb \cite{lhcb19} experiments and the
current experimental world average \cite{wa14} is
\bea
{\overline{\rm BR}}(B_s \to \mu^+ \mu^-)=\left (2.9 \pm 0.7 \right ) \times 10^{-9},~~~~~~{\overline{\rm BR}}(B_d \to \mu^+ \mu^-)
=\left (3.6_{-1.4}^{+1.6} \right ) \times 10^{-10}\;,
\eea
which are more or less consistent with the latest SM prediction (\ref{brmu}), but certainly they
do not rule out the possibility of new physics as the experimental errors are still quite large.

We will now consider the additional contributions arising due to the effect of scalar leptoquarks in this mode. 
Including the contributions arising from scalar
leptoquark exchange, one can write the transition amplitude for this
process from Eq. (\ref{hammu}) as
\bea
{\cal M}(B_q^0 \to \mu^+ \mu^-)=\langle \mu^+ \mu^- |{\cal H}_{eff}| B_q^0 \rangle
= - \frac{G_F}{ {\sqrt 2}~ \pi}V_{tb}V_{tq}^* \alpha f_{B_q} M_{B_q} m_\mu C_{10}^{SM} P ,
\eea
where
\be
P \equiv \frac{C_{10}-C_{10}'}{C_{10}^{SM}}=1+ \frac{C_{10}^{NP}-C_{10}^{'NP}}{C_{10}^{SM}}=1+r e^{i \phi^{NP}}\;,
\ee
with
\be
r e^{i \phi^{NP}}= (C_{10}^{NP}-C_{10}^{'NP})/C_{10}^{SM}\;,\label{bound}
\ee
$r$ denotes the magnitude of the ratio of  NP to SM  contributions and $\phi^{NP}$ is the relative phase between them. 
 As discussed in section II,  the exchange of the leptoquarks $X(3,2,7/6)$ and  $X(3,2,1/6)$  give new 
additional contributions to the Wilson coefficients  $C_{10}$ and  $C_{10}'$ respectively. Thus, the branching ratio
in both the cases will be
\be
{\overline{\rm BR}}(B_q \to \mu^+ \mu^-) = {\overline{\rm BR}}(B_q \to \mu^+ \mu^-)|_{\rm SM}(1+r^2 -2 r \cos \phi^{NP})\;.
\ee
Using the  theoretical and  experimental  branching ratios from (13) and (14),  the constraints on the combination of LQ couplings
can be obtained by requiring that  each individual leptoquark contribution  to the branching ratio does not exceed
the experimental result. The allowed region
in $r-\phi^{NP}$ plane which are compatible with the $1-\sigma$ range of
the experimental data are shown in Fig.-1 for $B_d \to \mu^+\mu^-$ (left panel) and
for $B_s \to \mu^+ \mu^-$ (right panel).
\begin{figure}[htb]
\includegraphics[width=8.0 cm,height=6.5cm, clip]{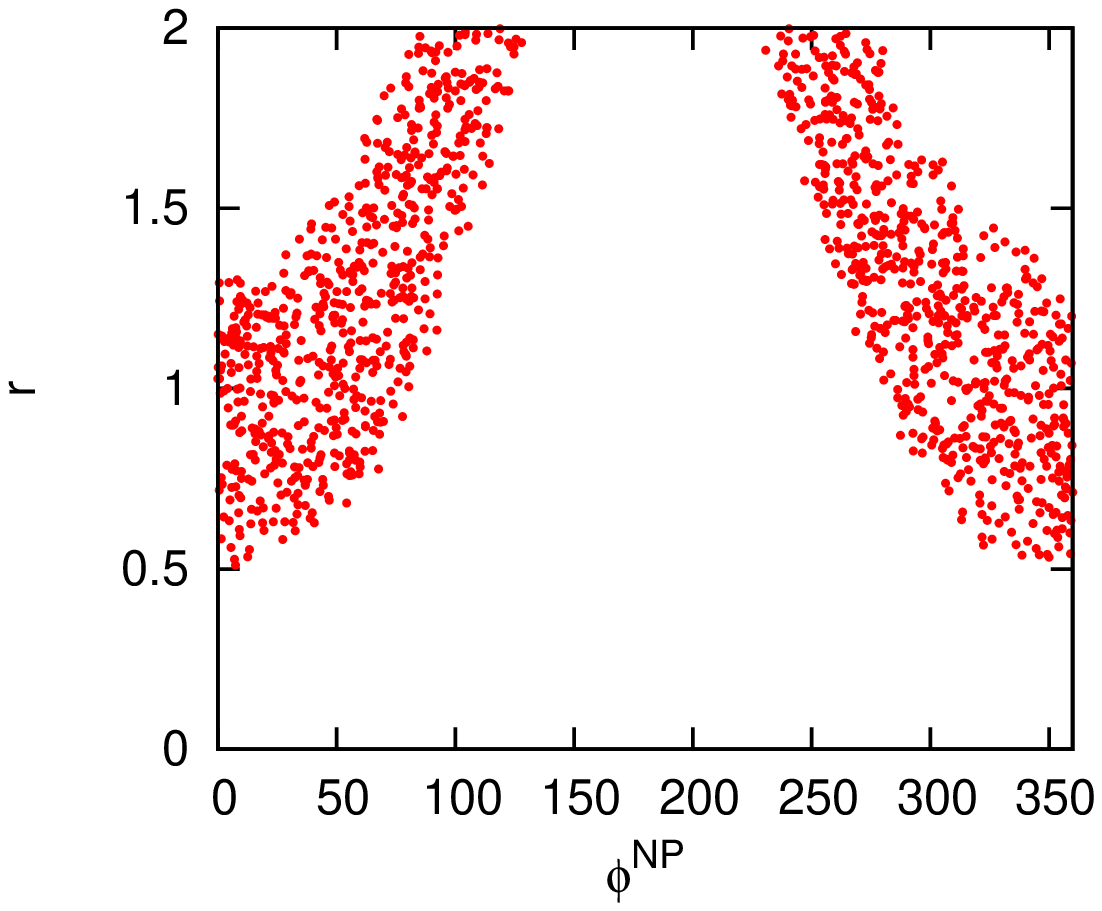}
\hspace{0.2 cm}
\includegraphics[width=8.0 cm,height=6.5cm, clip]{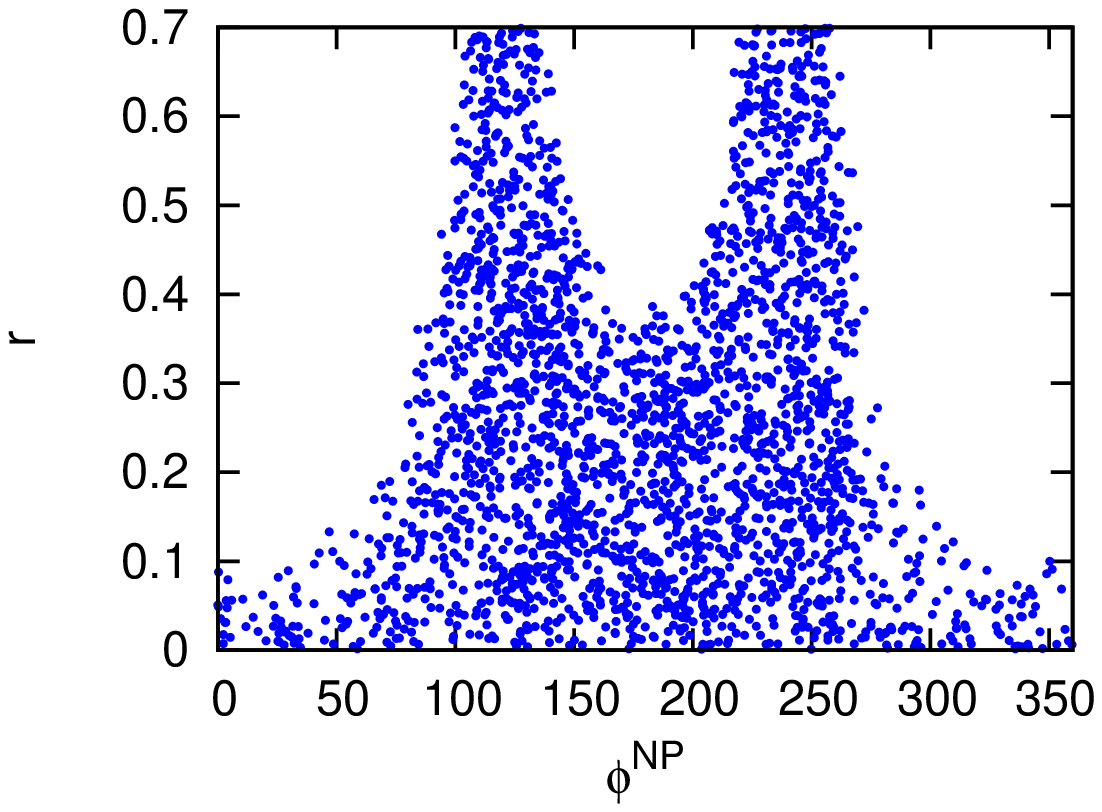}

\caption{The allowed region in the $r-\phi^{NP}$ parameters space obtained from the
${\rm BR}(B_d \to \mu^+ \mu^-)$ (left panel) and ${\rm BR}(B_s \to \mu^+ \mu^-)$ (right panel).}
\end{figure}
From the figure one can see that for $B_d \to \mu^+\mu^-$ the allowed range of $r$ and
$\phi^{NP}$ as 
\bea
 0.5\leq r \leq 1.3\;, ~~~{\rm for}~~~\left (0 \leq \phi^{NP} \leq  \pi/2\right )~~{\rm or}~~\left (3 \pi/2 \leq \phi^{NP} \leq  2\pi 
\right ),\label{r-bnd}
 \eea
 which can be translated to obtain the bounds for the leptoquark couplings using Eqs. (\ref{c10np}), (\ref{c10np1}) and (\ref{bound}) as
 \bea
 1.5 \times 10^{-9} ~{\rm GeV^{-2}} \leq \frac{|\lambda^{32} {\lambda^{12}}^*|}{M_S^2} \leq 3.9 \times 10^{-9}~ {\rm GeV^{-2}} \;,
 \eea
 with $M_S$ as the leptoquark mass.
For $B_s \to \mu^+ \mu^-$ process for $0\leq r \leq 0.1 $ the entire range for $\phi^{NP}$ is allowed, i.e.,
\bea
 0\leq r \leq 0.1\;, ~~~~{\rm for}~~~~0 \leq \phi^{NP} \leq 2 \pi \;.
 \eea
However, in our analysis we will use relatively mild constraint as
 \bea
 0\leq r \leq 0.35\;, ~~~~{\rm with}~~~~\pi/2 \leq \phi^{NP} \leq 3 \pi/2\;.\label{r-bound1}
 \eea
 This gives the constraint on leptoquark couplings as
 \bea
 0 \leq \frac{|\lambda^{32} {\lambda^{22}}^*|}{M_S^2} \leq 5 \times 10^{-9}~ {\rm GeV^{-2}} ~~~~{\rm for}~~~~\pi/2 \leq \phi^{NP} \leq 3 \pi/2\;.
 \eea
 One can also obtain the constrain on the leptoquark couplings $| \lambda^{32} {\lambda^{22}}^*|/M_S^2$ from the inclusive
 measurements ${\rm BR}(\bar B_d^0 \to X_s \mu^+ \mu^-)$. However, as shown in Ref. \cite{rm1}, these constraints are more  
 relaxed than those obtained from $B_s \to \mu^+ \mu^-$. So in our analysis we will use the constraints obtained from 
$B_s \to \mu^+ \mu^-$.

For other leptonic decay channels i.e., $B_{s,d} \to e^+ e^-,~\tau^+ \tau^-$ only the experimental upper limits exists \cite{pdg}. Now
using  the theoretical predictions for these branching ratios from \cite{bsll}, we obtain the constrain on the upper limits of the various
combinations of leptoquark couplings as presented in Table-II. However, the constraints obtained from such processes are rather weak.

\begin{table}[htb]
\begin{center}
\vspace*{0.1 true in}
\begin{tabular}{|c|c|c|}
\hline
~Decay Process~ & ~Couplings involved~ & ~Upper bound of the~ \\
 & & ~couplings $\left (\rm{in~ GeV^{-2}} \right )$ ~ \\
\hline
$B_d \to e^\pm e^\mp $ &~ $\frac{|\lambda^{31} {\lambda^{11}}^*|}{M_S^2}$ ~& ~$ < 1.73 \times 10^{-5}  $~\\

\hline

$B_d \to \tau^\pm \tau^\mp $ &~ $\frac{|\lambda^{33} {\lambda^{13}}^*|}{M_S^2}$ ~& ~$ < 1.28 \times 10^{-6} $~\\

\hline

$B_s \to e^\pm e^\mp $ &~ $\frac{|\lambda^{31} {\lambda^{21}}^*|}{M_S^2}$ ~& ~$ < 2.54 \times 10^{-5} $~\\

\hline

$B_s \to \tau^\pm \tau^\mp $ &~ $\frac{|\lambda^{33} {\lambda^{23}}^*|}{M_S^2}$ ~& ~$ < 1.2 \times 10^{-8} $~\\

\hline
\end{tabular}
\end{center}
\caption{Constraints obtained from the leptoquark couplings from various leptonic $B_{s,d} \to l^+ l^-$ decays.}
\end{table}

For the analysis of $B \to K e^+ e^-$ process, we need to know the values of the leptoquark couplings $\lambda^{31} {\lambda^{21}}^*/M_S^2$,
which can be extracted from the inclusive decay rates $B \to X_s e^+ e^-$. To obtain such constraints, we closely follow the procedure 
adopted in Ref. \cite{rm1}. Using the SM predictions and the corresponding experimental measurements from \cite{inclusive} for both
low-$q^2$ $(1-6)~{\rm GeV}^2$ and high-$q^2$ $(\gtrsim14.2 ~{\rm GeV}^2$) as
\bea
{\rm BR}(B \to X_s ee)|_{q^2 \in [1,6]~{\rm GeV}^2}&=& (1.73 \pm 0.12)\times 10^{-6}~{\rm (SM~prediction)}\nn\\
&=& (1.93 \pm 0.55)\times 10^{-6}~{\rm (Expt.)}\nn\\
{\rm BR}(B \to X_s ee)|_{q^2 >14.2~{\rm GeV}^2}&=& (0.2 \pm 0.06)\times 10^{-6}~{\rm (SM~prediction)}\nn\\
&=& (0.56 \pm 0.19)\times 10^{-6}~{\rm (Expt.)}
\eea
\begin{figure}[htb]
\includegraphics[width=7.0 cm,height=6.5cm, clip]{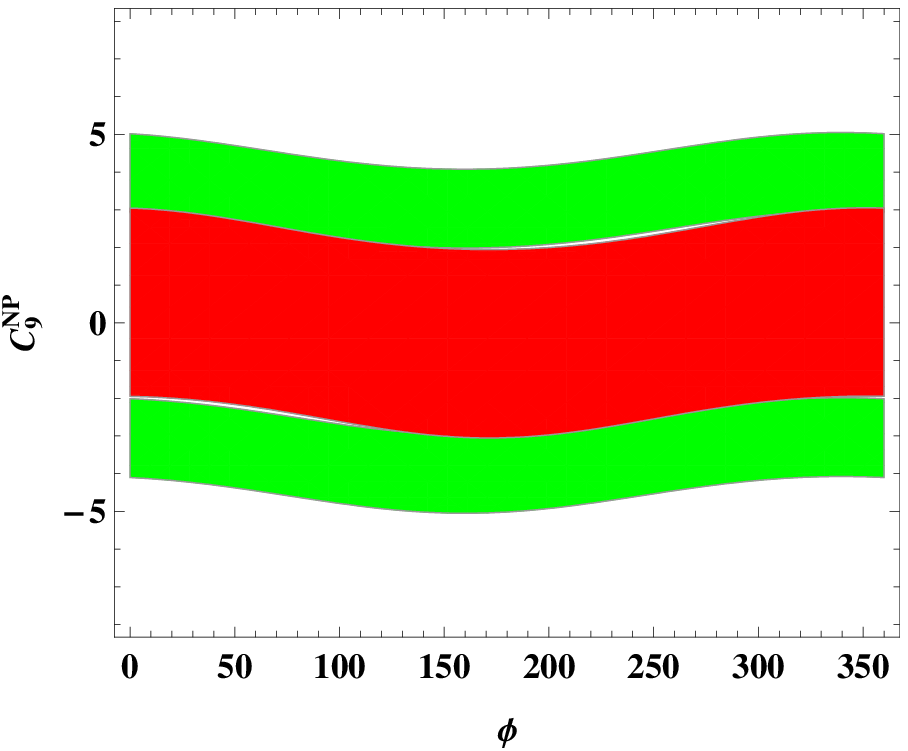}
\hspace{0.2 cm}
\includegraphics[width=7.0 cm,height=6.5cm, clip]{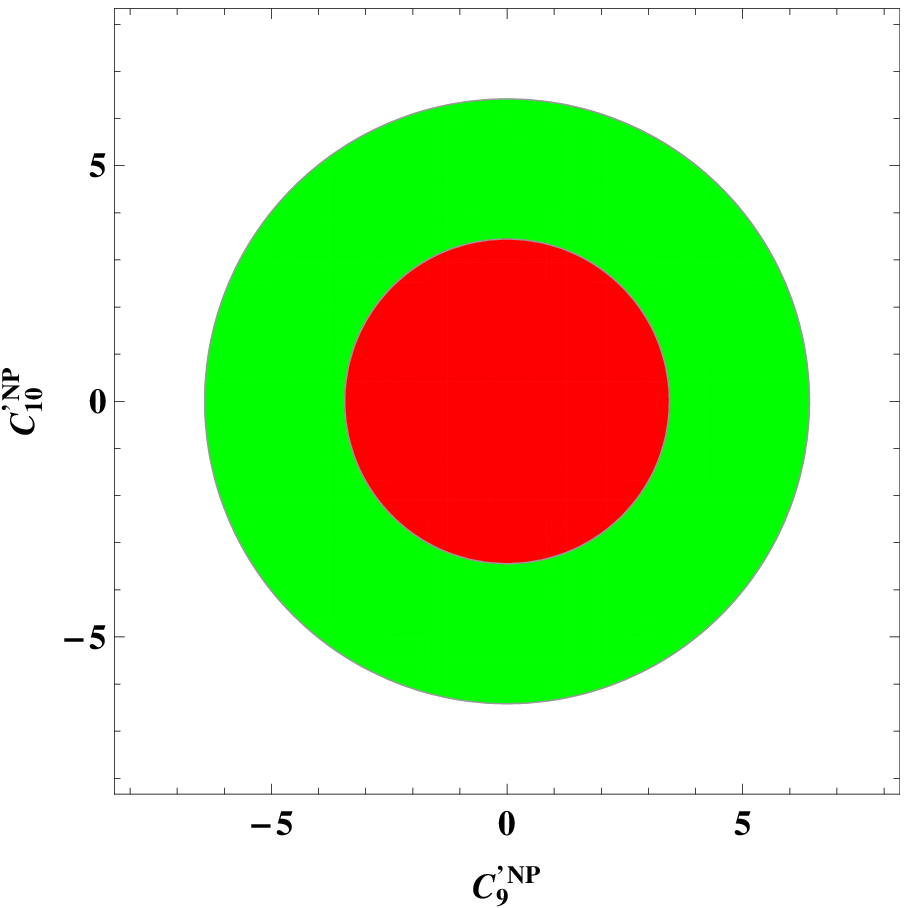}
\caption{The allowed region in $C_{10}^{NP}-\phi^{NP}$ parameter space (left panel) and 
$C_9^{'NP}-C_{10}^{'NP}$ (right panel)  obtained from 
$BR(B \to X_s e^+e^-)$, where the green (red) region corresponds to high-$q^2$ (low-$q^2$) limits.}
\end{figure}
In Fig-2, we show the allowed region in $C_{10}^{NP}-\phi^{NP}$ parameter space due to exchange of the leptoquark
$X(3,2,7/6)$ in the left panel.  The right panel depicts the allowed region in  the $C_9^{'NP}-C_{10}^{'NP}$  space
due to exchange of the leptoquark $X(3,2,1/6)$, where  green (red) region corresponds to high-$q^2$ (low-$q^2$) limits
in both the panels. Thus, it can be noticed that the bounds coming from the high-$q^2$ measurements are rather weak.
Considering the exchange of the $X(3,2,7/6)$ leptoquark as an example, we obtain the bound on $C_{10}^{NP}$ as
$-2.0 \leq C_{10}^{NP} \leq 3.0 $ for the entire range of $\phi^{NP}$, which gives the bound on $r$ as 
\be
0\leq r \leq 0.7\;.
\ee    
After obtaining the bounds on various leptoquark couplings, we now proceed to study the rare decays $B \to K/\pi l^+l^- $
and $B \to l_i^+ l_j^-$.

\section{$ B  \to K l^+ l^-$  process}

We now consider the  semileptonic decay process $B \to K l^+ l^-$, which is mediated by the quark level transition
$b \to s l^+ l^-$ and hence, it constitutes a quite
suitable tool of looking for new physics. The isospin asymmetries of $B \to K\mu^+ \mu^-$ and the partial branching ratios of
the decays $B^0 \to K^0 \mu^+ \mu^-$  and $B^+ \to K^+ \mu^+ \mu^-$ are
recently measured as functions of the dimuon mass squared ($q^2$) by the LHCb collaboration \cite{lhcb31}.
In this paper we will study the process in the large recoil region i.e., $1 \leq q^2 \leq 6~{\rm GeV}^2$,
in order to be well below the radiative tail of the charmonium resonances,  using the QCD factorization approach \cite{beneke,beneke1,beneke2}.
LHCb has measured the branching ratio in this region and the updated result is \cite{lhcb31}
\bea
{\rm BR}(B^+ \to K^+ \mu^+ \mu^-)|_{q^2 \in [1,6]~{\rm GeV}^2} = \left (1.19 \pm 0.03 \pm 0.06 \right ) \times 10^{-7}\;.
\eea
This mode has also been  analyzed by various  authors \cite{kll,kll-1,kll-2}  and the SM predictions is given as
\bea
{\rm BR}(B^+ \to K^+ \mu^+ \mu^-)|_{q^2 \in [1,6]~{\rm GeV}^2}^{SM} = \left (1.75 _{-0.29}^{+0.60} \right ) \times 10^{-7}\;.\label{sm-res}
\eea
Although, there is no significant discrepancy between these two results,  the SM predictions is slightly higher than
the experimental measurement.

To calculate the branching ratio, one use the effective Hamiltonian presented in Eq. (\ref{ham}) and obtain the transition amplitude for this process.
The matrix elements of the various hadronic currents between the initial $B$ meson and the final $K$ meson
can be parameterized in terms of the form factors  $f_+$, $f_0$ and $f_T$ as \cite{ball,hiller}
\bea
\langle K(p_K)| \bar s \gamma_\mu b | \bar{B}(p_B) \rangle =(2 p_B - q)_\mu f_+(q^2) + \frac{M_B^2-M_K^2}{q^2}q_\mu [f_0(q^2)-f_+(q^2)]
\eea
\bea
\langle K(p_K)| \bar s i \sigma_{\mu \nu} q^\nu b |\bar{B}(p_B) \rangle = -[(2 p_B -q)_\mu q^2 - (M_B^2 -M_K^2) q_\mu ] \frac{f_T(q^2)}{M_B+M_K}\;,
\eea
where the 4-momenta of the initial $B$-meson and the final  kaon are denoted by $p_B$ and $p_K$ respectively,  $M_B$, $M_K$ are the
corresponding masses and $q^2$ is the momentum transfer.
In the  large recoil region, the energy  of the kaon $E_K$ is large compared to
the typical size of hadronic binding energies ($\Lambda_{QCD}$) and the dilepton invariant mass
squared $q^2 $ is low. As a consequence, in this region the virtual photon
exchange between the hadronic part and the dilepton pair and the hard gluon scattering can
be treated in an expansion in $1/E_K$ \cite{hiller}, using either QCD factorization \cite{beneke1,beneke2} or Soft Collinear Effective Theory
(SCET) \cite{scet}. At the leading order in $1/E_K$ expansion, all the form factors $f_{+,0,T}(q^2)$ can be related to a single form factor $\xi_P(q^2)$.
Within QCDF approach, the form factor $f_+(q^2)$ is chosen to be $f_+(q^2) = \xi_P(q^2)$ and  including subleading corrections, the
other form factors can written as \cite{hiller}
\bea
\frac{f_0}{f_+} &=& \frac{2E_K}{M_B} \left [1+{\cal O}(\alpha_s)+{\cal O}\left (\frac{q^2}{M_B^2} \sqrt{\frac{\Lambda_{QCD}}
{E_K}}\right ) \right ]\nn\\
\frac{f_T}{f_+} &=& \frac{M_B+M_K}{M_B}\left[1+{\cal O}(\alpha_s)+{\cal O}\left (\sqrt{\frac{\Lambda_{QCD}}{E_K}} \right ) \right ]\;.
\eea
Thus,  only one soft form factor $\xi_P (q^2)$ appears in the $B \to K$
transition amplitude due to the symmetry relations in the large energy limit of QCD \cite{beneke, charles} and
the transition amplitude for the process $B \to K l^+l^-$  can be written as \cite{hiller}
\bea
{\cal M}(\bar B \to K \bar{l} l) &=& i\frac{G_F \alpha}{\sqrt 2 \pi} V_{tb}V_{ts}^* \xi_P(q^2)
\Big[F_V p_B^\mu (\bar l \gamma_\mu l)+ F_A p_B^\mu (\bar l \gamma_\mu \gamma_5 l)
+ F_P(\bar l  \gamma_5 l)\Big]\;,\label{ampbkll}
\eea
The functions $F_{V,A,P}(q^2)$ are given as
\bea
F_V & =& C_9 + \frac{2 m_b}{M_B} \frac{{\cal T}_P(q^2)}{\xi_P(q^2)}\nn\\
F_A &=& C_{10}\nn\\
F_P&=&  m_l C_{10} \Big[ \frac{M_B^2 -M_K^2}{q^2}\Big(\frac{f_0(q^2)}{f_+(q^2)}-1\Big) -1 \Big]\;.
\eea
The parameter ${\cal T}_P(q^2)$ takes into account the virtual one-photon exchange between the hadron
and the lepton pair and hard scattering contribution.  At lowest order,  it can be expressed as
\bea
{\cal T}_P^{(0)}(q^2)= \xi_P(q^2) \left ({C_7^{eff}}^{(0)}+ \frac{M_B}{2 m_b} Y^{(0)}(q^2) \right )\;.
\eea
The function $Y(q^2)$ denotes the perturbative part coming
from one loop matrix elements  of the four quark operators and
is given in Ref. \cite{beneke1}. The detailed expression for ${\cal T}_P(q^2)$, including  the subleading corrections
is presented in Appendix A.

With Eq. (\ref{ampbkll}), the double differential decay rate with respect to $q^2$ and $\cos \theta$ for the lepton
flavor $l$ is given as \cite{hiller}
 \bea
 \frac{d^2 \Gamma_l}{d q^2 d \cos \theta}= a_l(q^2)+c_l(q^2) \cos^2 \theta\;,
 \eea
 where
 \bea
 a_l(q^2)&=& \Gamma_0 \sqrt{\lambda} \beta_l \xi_P^2 \Big[q^2|F_P|^2+ \frac{\lambda}{4}
 (|F_A|^2+|F_V|^2)\nn\\
 &+& 2 m_l (M_B^2-M_K^2+q^2) Re(F_P F_A^*) +4 m_l^2 M_B^2 |F_A|^2 \Big]\;,\nn\\
  c_l(q^2)&=& -\Gamma_0 \sqrt{\lambda} \beta_l^3 \xi_P^2  \frac{\lambda}{4}
 \left (|F_A|^2+|F_V|^2\right ),
 \eea
\be
\lambda =M_B^4 +M_K^4+q^4 -2(M_B^2 M_K^2+M_B^2 q^2+M_K^2 q^2),~~~~~\beta_l = \sqrt{1-4 m_l^2/q^2}\;,
\ee
and
\bea
\Gamma_0 = \frac{G_F^2 \alpha^2 |V_{tb}V_{ts}^*|^2}{2^9 \pi^5 M_B^3}\;.
\eea
The decay rate can be expressed as
\be
\Gamma_l =2 \left (A_l + \frac{1}{3}C_l \right )\;,
\ee
where the $q^2$-integrated coefficients are given as
 \bea
 A_l = \int_{q_{min}^2}^{q_{max}^2} d q^2 a_l(q^2)\;,~~~~~~~~~~~~~~
 C_l = \int_{q_{min}^2}^{q_{max}^2} d q^2 c_l(q^2)\;.
 \eea
The observable   $R_K$ which is the ratio of $B \to K
\mu^+  \mu^-$ to $B \to K e^+e^-$ decay rates with same $q^2$ cuts is
\be
R_K\equiv \frac{\Gamma_\mu}{\Gamma_e} = \int_{q_{min}^2}^{q_{max}^2} dq^2 \frac{d \Gamma_\mu}{d q^2}{\biggr /}
\int_{q_{min}^2}^{q_{max}^2} dq^2 \frac{d \Gamma_e}{d q^2}\;.
\ee
which probes the lepton flavour non-universality effects.

Another observable, which can be constructed from ratios or asymmetries where the leading form factor uncertainties cancel.
The CP-averaged isospin asymmetry is such an observable which is defined as
\bea
A_I(q^2) &=& \frac{d\Gamma(B^0 \to K^0 \mu^+ \mu^-)/dq^2 - d\Gamma(B^+ \to K^+ \mu^+ \mu^-)/dq^2}
{d\Gamma(B^0 \to K^0 \mu^+ \mu^-)/d q^2 + d\Gamma(B^+ \to K^+ \mu^+ \mu^-)/dq^2}\nn\\
&=& \frac{d {\rm BR}(B^0 \to K^0 \mu^+ \mu^-)/d q^2 - (\tau_0/\tau_+) d{\rm BR}(B^+ \to K^+ \mu^+ \mu^-)/dq^2}
{d{\rm BR}(B^0 \to K^0 \mu^+ \mu^-)/dq^2 + (\tau_0/\tau_+)d{\rm BR}(B^+ \to K^+ \mu^+ \mu^-)/dq^2}\;.
\eea

With these formulae at hand, we now proceed for numerical estimation. To make predictions for SM observables, or to extract information
about potentially new short distance physics, one should require the knowledge
of associated hadronic  form factors.
For this purpose we use value of  the form factors $f_+(q^2) = \xi_P(q^2)$  calculated in the light-cone sum rule (LCSR)
approach \cite{ball}, where the $q^2$ dependence is given
by simple fits as
\be
f_+(q^2) = \frac{r_1}{1-s/m_{fit}^2} + \frac{r_2}{\left (1-s/m_{fit}^2\right )^2}\;,
\ee
and the values of the parameters used  are taken from \cite{ball}. 
The particle masses and the lifetime of $B_{s}$ meson, the decay constants  are taken from \cite{pdg} and the  SM Wilson coefficients $C_i$'s
are listed in Table-1. For the CKM matrix elements
we use the Wolfenstein parametrization with the values $A=0.814_{-0.024}^{+0.023}$, $\lambda=0.22537\pm 0.00061$,
$\bar{ \rho} =0.117 \pm 0.021$ and $\bar{ \eta} =0.353 \pm 0.013$ \cite{pdg}.
The values of the quark  masses used in our analysis are as follows.
The $b$-quark mass used in ${\cal T}_P$ and $F_V$ is the potential subtracted (PS) mass $m_{PS}(\mu_f)$
 at the factorization scale $\mu_f \sim \sqrt{\Lambda_{QCD} m_b}$ and is denoted by $m_b$. 
The function $Y (q^2)$ is evaluated by using the  pole mass $m_b^{pole}$ and its  relation to the
PS mass is given as $m_b^{pole}= m_b^{PS}(\mu_f)+4 \alpha_s \mu_f/3 \pi$.
The quark masses (in GeV) used are $m_b$=4.6, $m_c$=1.4, the fine structure coupling constant $\alpha=1/130$. 
Using these values we show in Fig.-3 the variation of differential branching ratio for
$B^0 \to K^0 \mu^+ \mu^-$ (left panel) and the $B^+ \to K^+ \mu^+ \mu^-$ (right panel)  in the standard model
with respect to the di-muon invariant mass. The variation of isospin asymmetry and  $R_K$ are shown in Fig- 4.
The total branching ratios integrated over the range $q^2\in[1,6]~{\rm GeV}^2$ are summarized in Table-3.
Our predictions for branching fractions are in agreement with the previous predictions \cite{kll}, and the slight difference can be attributed
to the difference in the values of input parameters used in the calculation. But these predictions are slightly larger
than the experimental values. The $q^2$-averaged isospin asymmetry, $\langle A_I(q^2) \rangle$ can be obtained from Eq. (41)
by replacing the differential branching fractions with the corresponding integrated values.

\begin{figure}[htb]
\includegraphics[width=8.0 cm,height=6.5cm, clip]{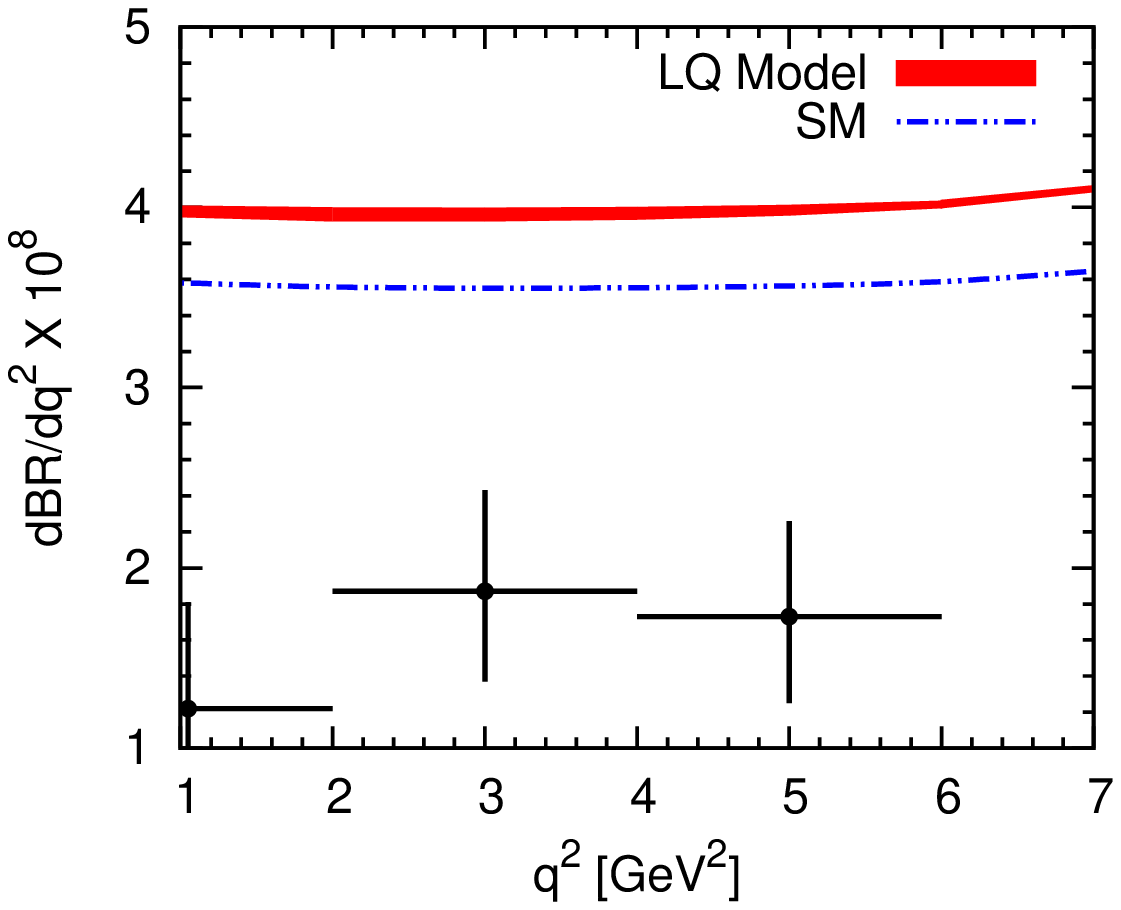}
\hspace{0.2 cm}
\includegraphics[width=8.0 cm,height=6.5cm, clip]{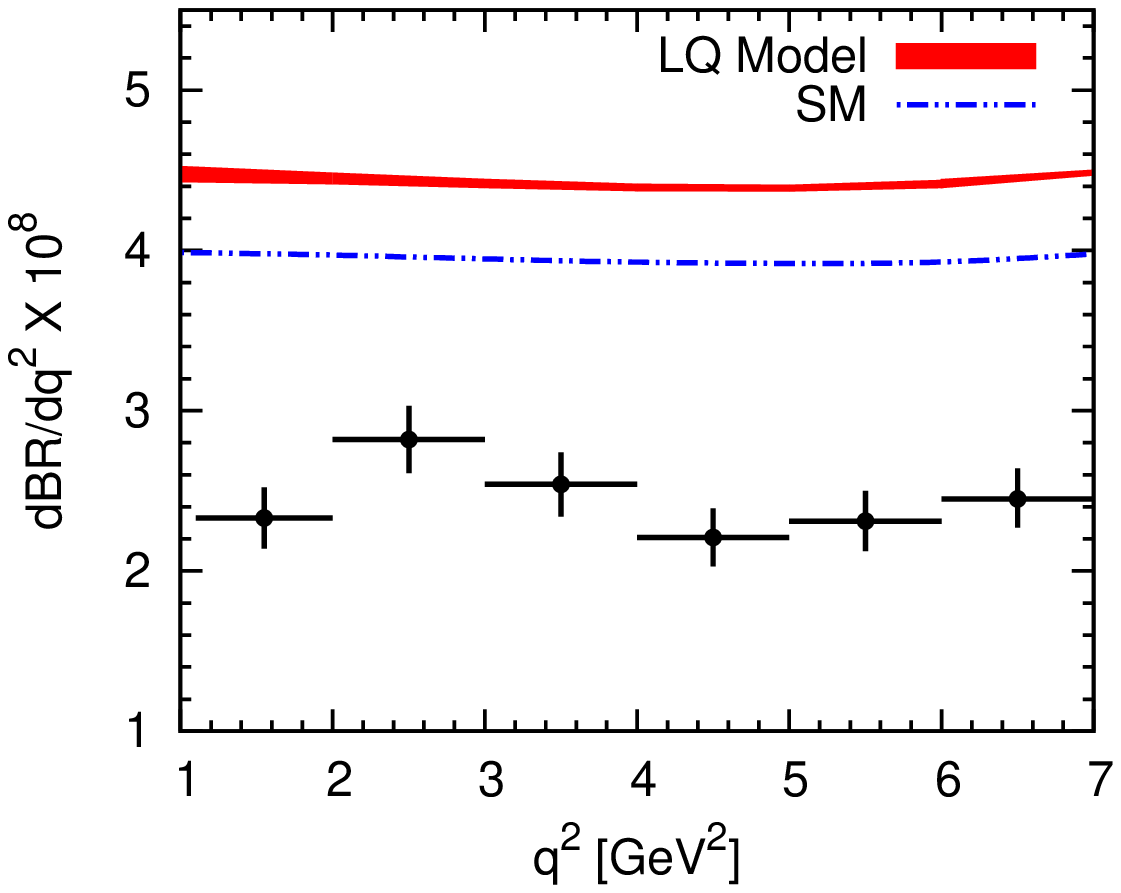}
\caption{The variation of branching ratios with $q^2$ for the decay processes $B^0 \to K^0 \mu^+ \mu^-$ (left panel)
and $B^+ \to K^+ \mu^+ \mu^-$ (right panel) in standard model and in leptoquark model. The red band for the leptoquark model is obtained
by using the allowed leptoquark parameter space. The $q^2$-averaged (bin-wise) 1-$\sigma$ experimental results are shown by black plots, where
horizontal (vertical) line denotes the bin width (1-$\sigma$ error). }
\end{figure}

\begin{figure}[htb]
\includegraphics[width=8.0 cm,height=6.5cm, clip]{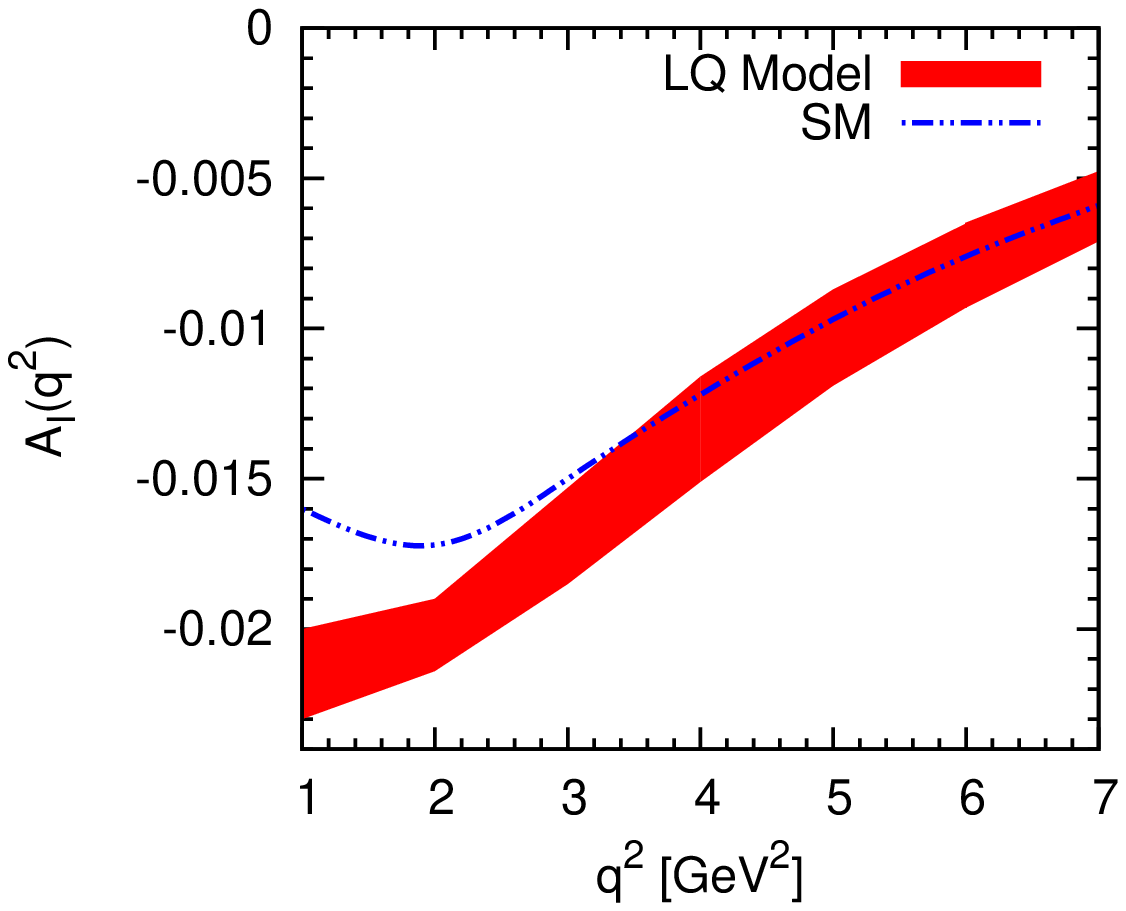}
\hspace{0.2 cm}
\includegraphics[width=8.0 cm,height=6.5cm, clip]{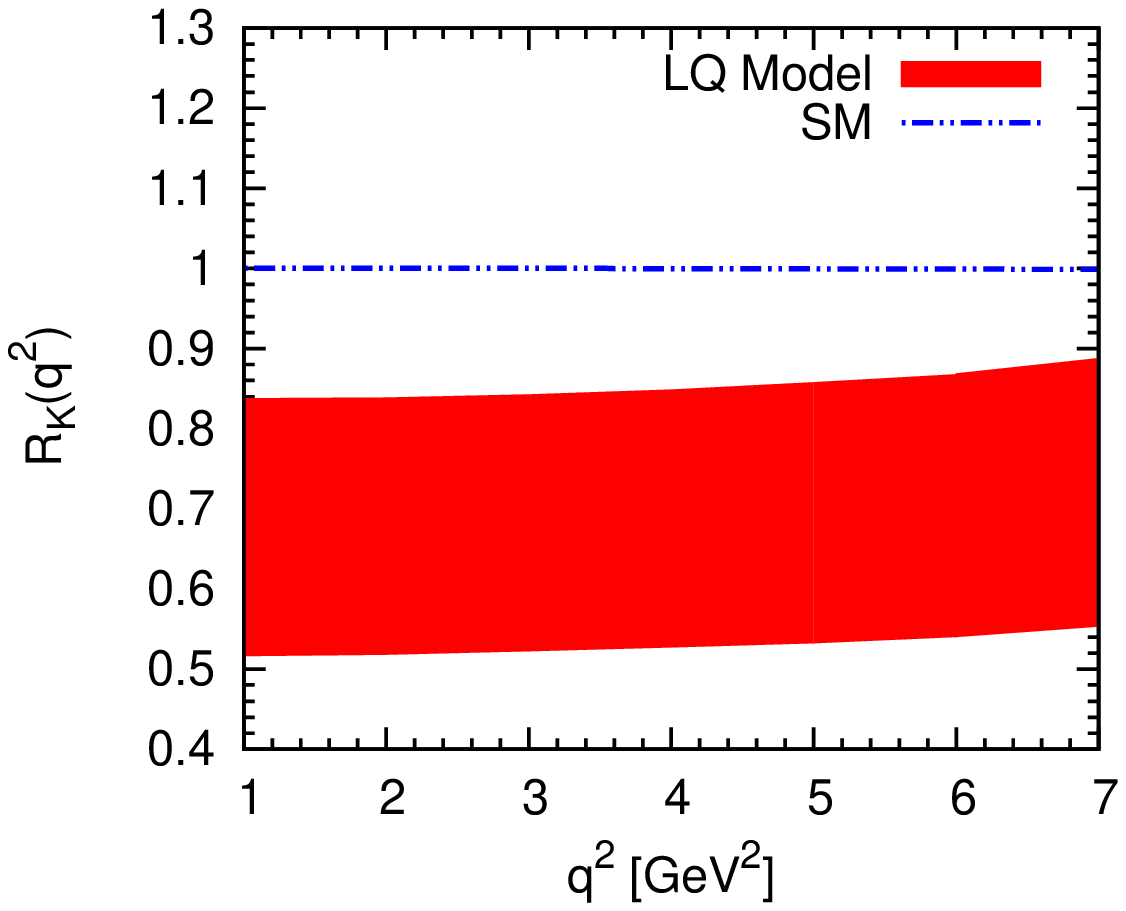}
\caption{The variation of isospin asymmetry (left panel) for $B \to K \mu \mu $ and $R_K$ (right panel) with $q^2$.}
\end{figure}

In the leptoquark model, these processes will receive additional contributions arising from the leptoquark
exchange and hence, the Wilson coefficients $C_{9,10}$ will receive additional contributions
$C_{9,10}^{NP}$  as well as
new Wilson $C_{9,10}'$ associated with the chirally flipped operators $O_{9,10}'$ will also be present
as already discussed in Section II. The bounds on these new Wilson coefficients can be obtained from
the constraint on $r$ (\ref{r-bound1}) which has been extracted from the
experimental results on BR$(B_s \to \mu^+ \mu^-)$. Thus, for the leptoquarks $X=(3,2,7/6)$ and $X=(3,2,1/6)$,
we obtain the value of $r\lesssim 0.35$ for $\pi/2 \leq \phi^{NP}\leq 3 \pi/2  $, which  can be
translated with eqns (7), (10) and (17) to give the value of the new Wilson coefficients as
\bea
&&|C_9^{LQ}|= |C_{10}^{LQ}|\leq |r ~C_{10}^{SM}| ~~~~~~~~({\rm for}~X=(3,2,7/6))\nn\\
&&|C_9^{'~LQ}|=|C_{10}^{'~LQ}| \leq |r~C_{10}^{SM}|~~~~~~({\rm for}~X=(3,2,1/6))\;.
\eea

\begin{table}[htb]
\begin{center}
\vspace*{0.1 true in}
\begin{tabular}{|c|c|c|c|}
\hline
Observables & SM Predictions & Values in LQ model   \\
\hline
$B_d \to K^0 \mu^+ \mu^- $ & $1.82 \times 10^{-7}$ &~$(2.04 - 2.16 )\times 10^{-7}$ ~\\
$B^+ \to K^+ \mu^+ \mu^- $ & $1.99 \times 10^{-7}$& $(2.2 - 2.3 )\times 10^{-7}$\\
$B^+ \to K^+ e^+ e^- $ & $1.82 \times 10^{-7}$& $(2.3 - 3.7 )\times 10^{-7}$\\
$\langle  A_I\rangle $ & $-0.03$ & $-0.036 \to -0.024$ \\
$  R_K $ & 1.09 & $0.62 - 0.96 $ \\
\hline
\end{tabular}
\end{center}
\caption{The predicted  values for the integrated branching ratio and isospin asymmetry in the range
$q^2 \in[1,6]~{\rm GeV}^2$ for the decay mode
$B \to K \mu \mu $ and the value of $R_K$ in the SM as well as in leptoquark model. The red band is obtained
by varying the LQ parameters within their allowed ranges. }
\end{table}

Using these values we show the variation of differential branching ratio and isospin asymmetry
and $R_K$ for $X=(3,2,7/6)$ in Figs.-3 and 4. For the calculation of the $B^+ \to K^+ e^+ e^-$
in the determination of $R_K$, we have used the constraint on the leptoquark couplings obtained from $B \to X_s ee$
inclusive decay rate.  From these figures it can be
seen that there is slight deviation in $B \to K \mu \mu$  branching ratios  from their SM values. The isospin asymmetry 
also has slight deviation from its SM prediction and this deviation is substantial in the low -$q^2$ region.
However, the $R_K$ value deviates significantly and it is possible to accommodate the observed experimental value
in the leptoquark model. The integrated branching ratios and the isospin asymmetries are presented in Table-III. 
 
\section{$B \to \pi ll $ Process}
In this section we would like to study the  decay mode $B \to \pi \mu^+ \mu^-$ which is mediated by the quark level transition
$b \to d l^+ l^-$. This decay mode has been recently observed by the LHCb \cite{lhcb15} collaboration and the measured branching
ratio is
\bea
{\rm BR}(B^+ \to \pi^+ \mu^+ \mu^-) = \left (2.3 \pm 0.6~ ({\rm stat}) \pm 0.1~ ({\rm syst}) \right ) \times 10^{-8}\;,
\eea
at $5.2 \sigma$ significance.

For the calculation of branching ratio, we closely follow \cite{hou} and here we preview only the main results.
In the standard model the effective Hamiltonian for $b \to d l^+ l^-$ transition is given by 
\bea
{\cal H}_{eff} = - \frac{G_F}{\sqrt 2} \left [\lambda_t^{(d)}{\cal H}_{eff}^{(t)}
+ \lambda_u^{(d)}{\cal H}_{eff}^{(u) }\right ]+ h.c.
\eea
where $\lambda_q^{(d)}=V_{qb}V_{qd}^*$ and
\bea
{\cal H}_{eff}^{(u)}=C_1(O_1^c-O_1^u)+C_u(O_2^c -O_2^u)\nn\\
{\cal H}_{eff}^{(t)}=C_1 O_1^c + C_2 Q_2^c + \sum_{i=3}^{10} C_i O_i
\;.
\eea
It should be noted that for $b \to d l^+l^-$ transitions  $\lambda_u^{(d)}$ and $\lambda_t^{(d)}$ are
comparable in magnitude  with the phase difference $\phi_2 = {\rm arg} \left(-V_{td}V_{tb}^*/
V_{ud}V_{ub}^* \right )$. Thus, one can write the transition amplitude for the process $B \to \pi l^+ l^- $ as
\bea
{\cal M}(B(p) \to \pi(p') l^+l^-)& =& \frac{G_F \alpha }{2 \sqrt 2 \pi}c_\pi^{-1} \xi_\pi \Big[\lambda_t C_{9,\pi}^{(t)}+ \lambda_u C_{9,\pi}^{(u)}(p+p')^\mu
(\bar l \gamma_\mu l) \nn\\
&+& \lambda_t C_{10} (p+p')^\mu (\bar l \gamma_\mu \gamma_5 l) \Big ]\;,
\eea
where  $c_\pi=1/\sqrt{2}$ for $\pi^0$ and 1 for $\pi^\pm$ and
\bea
C_{9,\pi}^{(t)}(q^2) &=& C_9 + \frac{2 m_b}{M_B} \frac{{\cal T}_\pi^{(t)}(q^2)}{\xi_\pi(q^2)}\nn\\
C_{9,\pi}^{(u)}(q^2) & =& \frac{2 m_b}{M_B} \frac{{\cal T}_\pi^{(u)}(q^2)}{\xi_\pi(q^2)}\;.
\eea
The differential branching ratio is given as
\bea
\frac{d {\rm BR}}{d q^2} (B \to \pi l^+ l^-) &=& S_\pi\tau_B \frac{G_F^2 M_B^3}{96 \pi^3} \left (\frac{\alpha}{4 \pi} \right )^2 \lambda_\pi^3
\xi_\pi(q^2)^2 |\lambda_t|^2\nn\\
&\times & \Big (\left|C_{9, \pi}^{(t)} (q^2) - R_{ut} e^{i \phi_2}C_{9, \pi}^{(u)}(q^2)\right |^2 + |C_{10}|^2 \Big )
\eea
where
\be
\lambda_\pi(q^2, m_\pi^2)= \left[ \left (1- \frac{q^2}{M_B^2} \right )^2 - \frac{2 m_\pi^2}{M_B^2} \left (1+ \frac{q^2}{M_B^2} \right )
+ \frac{m_\pi^4}{M_B^4} \right ]^{1/2}\;.
\ee
with $S_\pi=1/c_\pi^2$
and $\lambda_u^{(d)}/\lambda_t^{(d)} = -R_{ut} e^{i \phi_2}$.
The branching ratio for the CP conjugate mode can be obtained by changing the sign of the weak phase $\phi_2$. One can then define the $q^2$
dependence of the direct CP asymmetries as
\bea
A_{CP}^{+}(q^2) &=& \frac{d {\rm BR}(B^- \to \pi^- ll)/d q^2- d {\rm BR}(B^+ \to \pi^+ ll)/d q^2}
{d {\rm BR}(B^- \to \pi^- ll)/d q^2+ d {\rm BR}(B^+ \to \pi^+ ll)/d q^2}\nn\\
A_{CP}^{0}(q^2) &=& \frac{d {\rm BR}(\bar{B}^0 \to \pi^0 ll)/d q^2- d {\rm BR}(B^0 \to \pi^0 ll)/d q^2}
{d {\rm BR}(\bar{B}^0 \to \pi^0 ll)/d q^2+ d {\rm BR}(B^0 \to \pi^0 ll)/d q^2}\;.
\eea
The $q^2$ dependent isospin asymmetry is defined as
\be
A_I(q^2) = \frac{\tau_{B^0}}{2 \tau_{B^\pm}} \frac{d \overline{\rm BR}(B^+ \to \pi^+ l^+l^-)/dq^2}{d \overline{\rm BR}(B^0 \to \pi^0 l^+l^-)/dq^2}-1.
\ee
where $\overline{\rm BR}$ is the CP averaged branching ratio.

The $B \to \pi$ form factor can be obtained using the light-cone QCD sum rule approach
\be
\xi_\pi(q^2) = \frac{\xi_\pi(0)}{(1-q^2/m_{B^*}^2)(1- \alpha_{BK}q^2/m_B^2)}\;,
\ee
where the numerical value for the normalization constant is $\xi_\pi(0)=0.26_{-0.03}^{+0.04}$ and the slope parameter
$\alpha_{BK} = 0.53 \pm 0.06$. The light cone distribution amplitude is given by
\be
\phi_\pi(u)= 6 u (1-u)\Big[1+a_2^\pi C_2^{(3/2)}(2u-1)+a_4^\pi C_4^{(3/2)}(2u-1)+ \cdots \Big]\;,
\ee
where $C_n^{3/2}(x)$ are Gegenbauer polynomials and the coefficients $a_i^\pi$ are related to the moments of distribution
amplitudes (DAs). The numerical values of these coefficients are
$a_2^\pi=0.25 \pm 0.15$,  $a_2^\pi + a_4^\pi = 0.1 \pm 0.1$ as given in Refs. \cite{hou,ball}.

The $B$ meson light cone distribution amplitudes can be given as
\bea
\Phi_{B,+}(\omega) = \frac{\omega}{\omega_0^2} e^{-\omega/\omega_0},~~~~~~~\Phi_{B,-}(\omega) = \frac{1}{\omega_0} e^{-\omega/\omega_0}
\eea
with $\omega_0 =2 \bar{\Lambda}_{HQET}/3$ and $ \bar{\Lambda}_{HQET}=m_B-m_b$. These enter only through the moments
\bea
\lambda_{B,+}^{-1} = \int_0^\infty d \omega \frac{\Phi_{B,+}(\omega)}{\omega} = \omega_0^{-1}\;,
\eea
\bea\lambda_{B,-}^{-1}(q^2) = \int_0^\infty d \omega \frac{\Phi_{B,-}(\omega)}{\omega - q^2/M_B -i \epsilon} =
\frac{e^{-q^2/M_B \omega_0}}{\omega_0} \Big[-Ei(q^2/M_B \omega_0) + i \pi \Big]\;,
\eea
where $Ei(z)$ is the exponential integral function. 
Using these formulae, we show in Fig. 5 the differential branching ratios for $\bar{B}^0 \to \pi^0 \mu^+ \mu^-$ (left panel) 
and $B^- \to \pi^- \mu^+ \mu^-$ (right panel) both in the SM and in leptoquark model, for which we have used the constraints 
on leptoquark couplings as extracted from $B_d \to \mu^+ \mu^-$ (\ref{r-bnd}). In this case the branching ratios in
the leptoquark model have significant deviations from their corresponding SM values. The integrated branching ratios
in the range $q^2 \in [1,6]GeV^2$ are presented in Table-IV. Similarly the variation of CP asymmetries are shown in Fig-6
and  the CP asymmetries averaged over $q^2$  range are given in Table-IV. The variation of isospin asymmetry is shown
in the left panel of Fig. 7.
\begin{table}[htb]
\begin{center}
\vspace*{0.1 true in}
\begin{tabular}{|c|c|c|c|}
\hline
Observables & SM Predictions & Values in LQ model   \\
\hline
$B_d \to \pi^0 \mu^+ \mu^- $ & ~$2.6 \times 10^{-9} $~ &$(3.2-3.4) \times 10^{-9} $ \\
$B^+ \to \pi^+ \mu^+ \mu^- $ & $5.6 \times 10^{-9} $&$(7.2-7.3) \times 10^{-9} $ \\

$\langle  A_{CP}^0\rangle $ & $-0.103$  & $-0.04 \to -0.065$ \\

$\langle  A_{CP}^+\rangle $ & $-0.268$ &$-0.11 \to -0.06$  \\
$\langle  A_I\rangle $ & 0.078 & 0.04 - 0.07 \\
\hline
\end{tabular}
\end{center}
\caption{The predicted  branching ratios, $q^2$ averaged CP asymmetries and the isospin asymmetry  for $B \to \pi \mu^+
\mu^-$ process.}
\end{table}

The ratio of branching ratios  of $B^+ \to \pi^+ \mu^+ \mu^-$ and $B^+ \to K^+ \mu^+ \mu^-$ has recently been measured by LHCb 
experiment \cite{lhcb15} as
\bea
\frac{{\rm BR}(B^+ \to \pi^+ \mu^+ \mu^-)}{{\rm BR}(B^+ \to K^+ \mu^+ \mu^-)}=0.053 \pm 0.014 ({\rm stat}) \pm 0.001~ ({\rm syst})
\eea
We define
\bea
R_+ (q^2)= \frac{d{\rm BR}(B^+ \to \pi^+ ll)/dq^2}{d{\rm BR}(B^+ \to K^+ ll)/dq^2}
\eea
and show the variation of $R_+(q^2)$ with dimuon invariant mass in the right panel of  Fig. 7

\begin{figure}[htb]
\includegraphics[width=8.0 cm,height=6.5cm, clip]{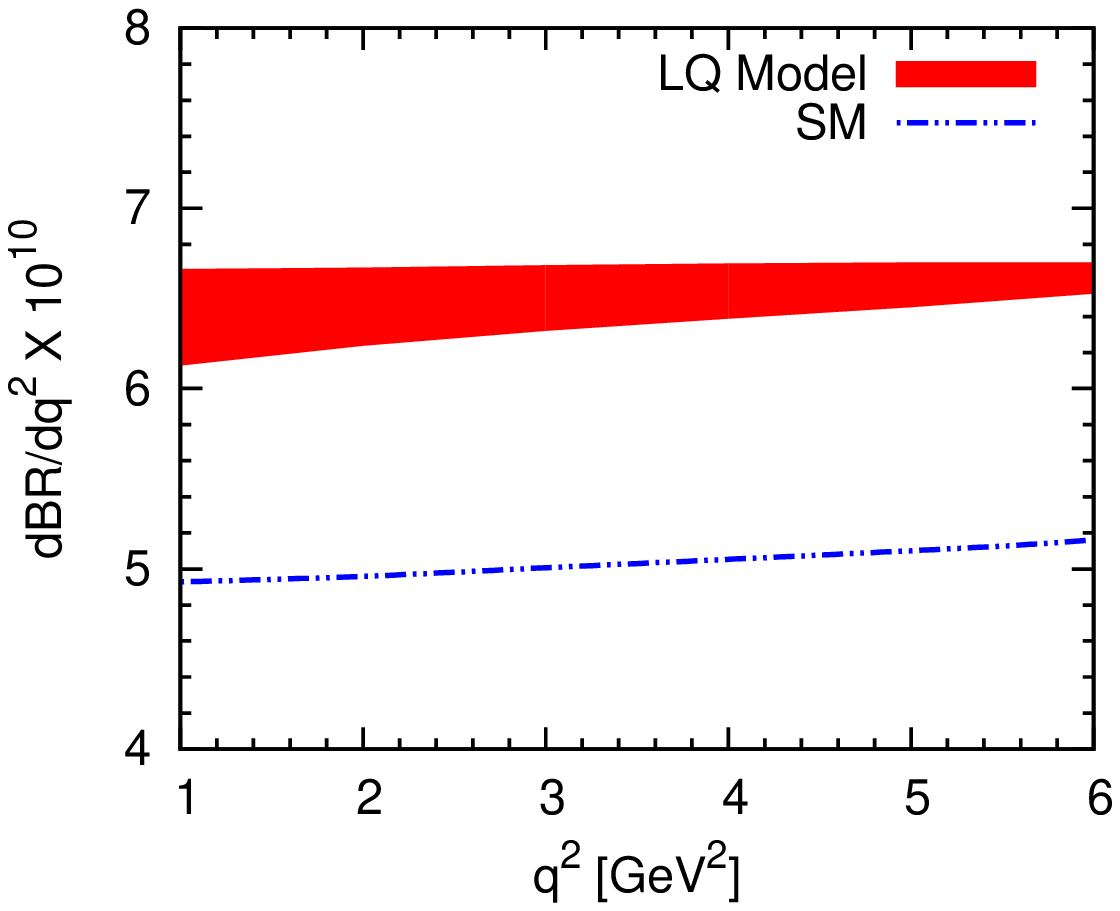}
\hspace{0.2 cm}
\includegraphics[width=8.0 cm,height=6.5cm, clip]{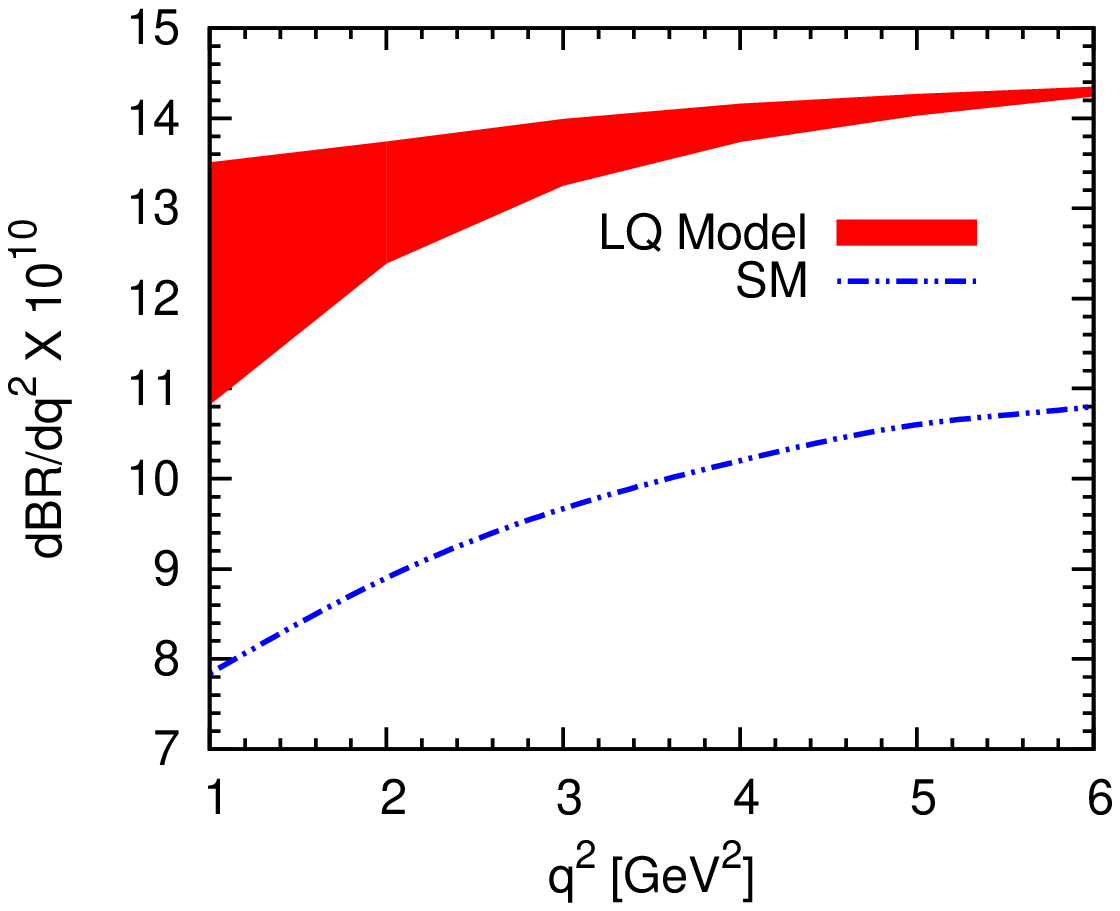}
\caption{The differential branching ratio for ${\bar B}^0 \to \pi^0 \mu^+ \mu^- $ (left panel) and $B^- \to \pi^- \mu^+ \mu^-$
(right panel) in the SM and lepto-quark model.}
\end{figure}

\begin{figure}[htb]
\includegraphics[width=8.0 cm,height=6.5cm, clip]{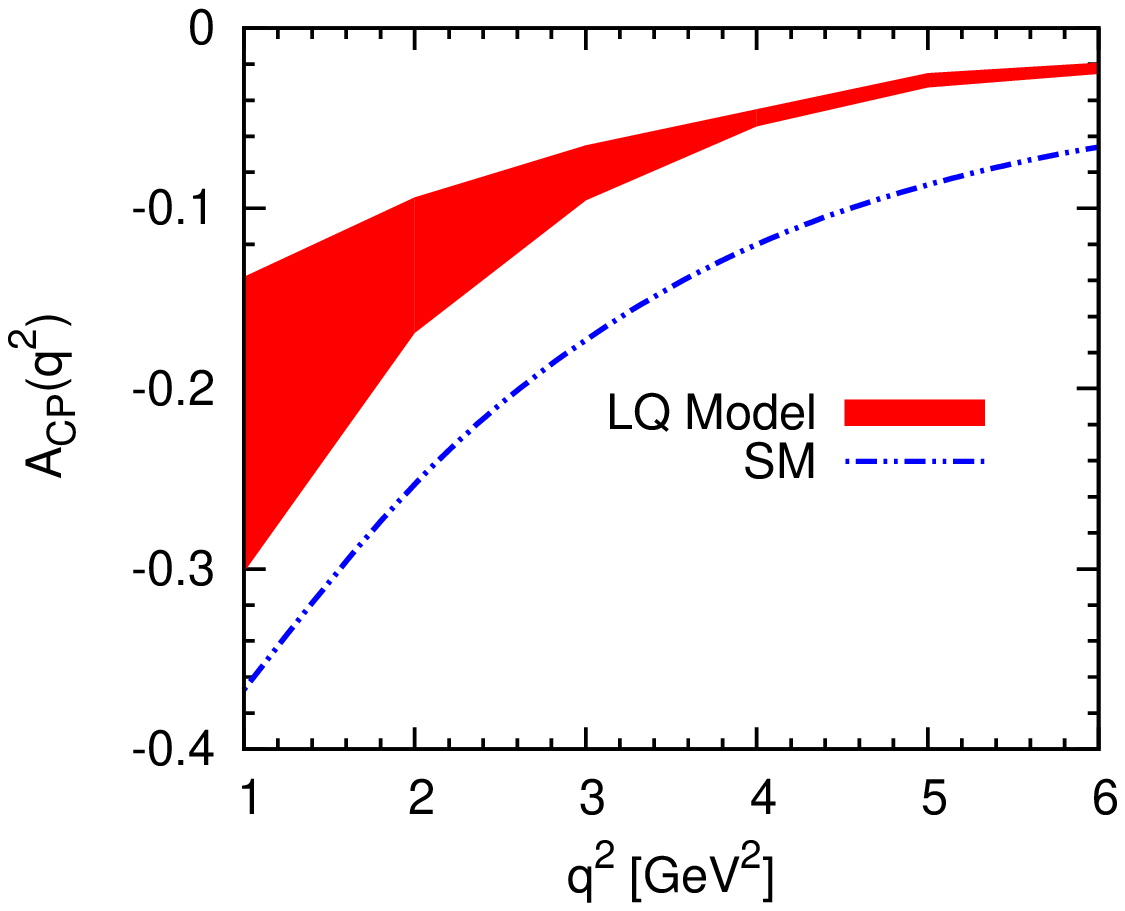}
\hspace{0.2 cm}
\includegraphics[width=8.0 cm,height=6.5cm, clip]{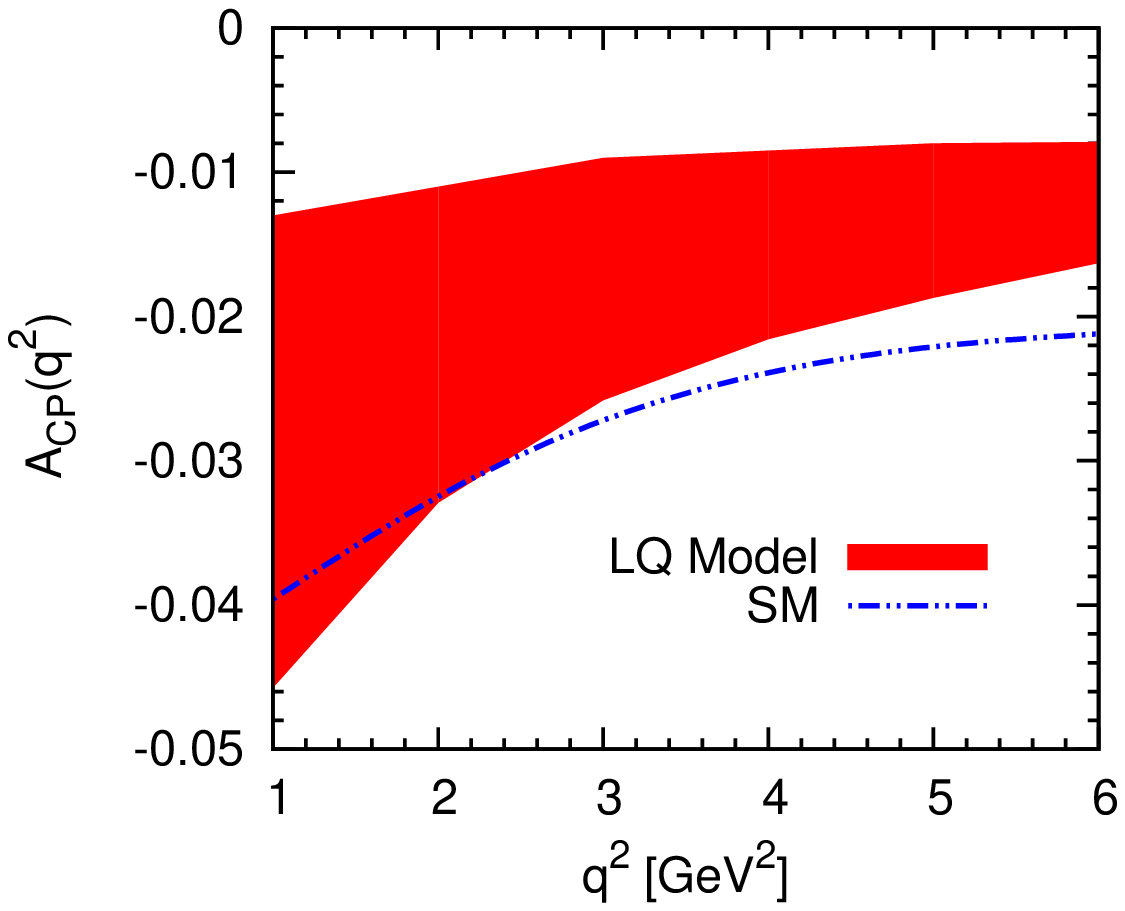}
\caption{Variation of $A_{CP}(q^2)$ for $B^0({\bar B}^0) \to \pi^0 \mu^+ \mu^- $ (left panel) and $B^\pm \to \pi^\pm \mu^+ \mu^-$ (right panel).}
\end{figure}

\begin{figure}[htb]
\includegraphics[width=8.0 cm,height=6.5cm, clip]{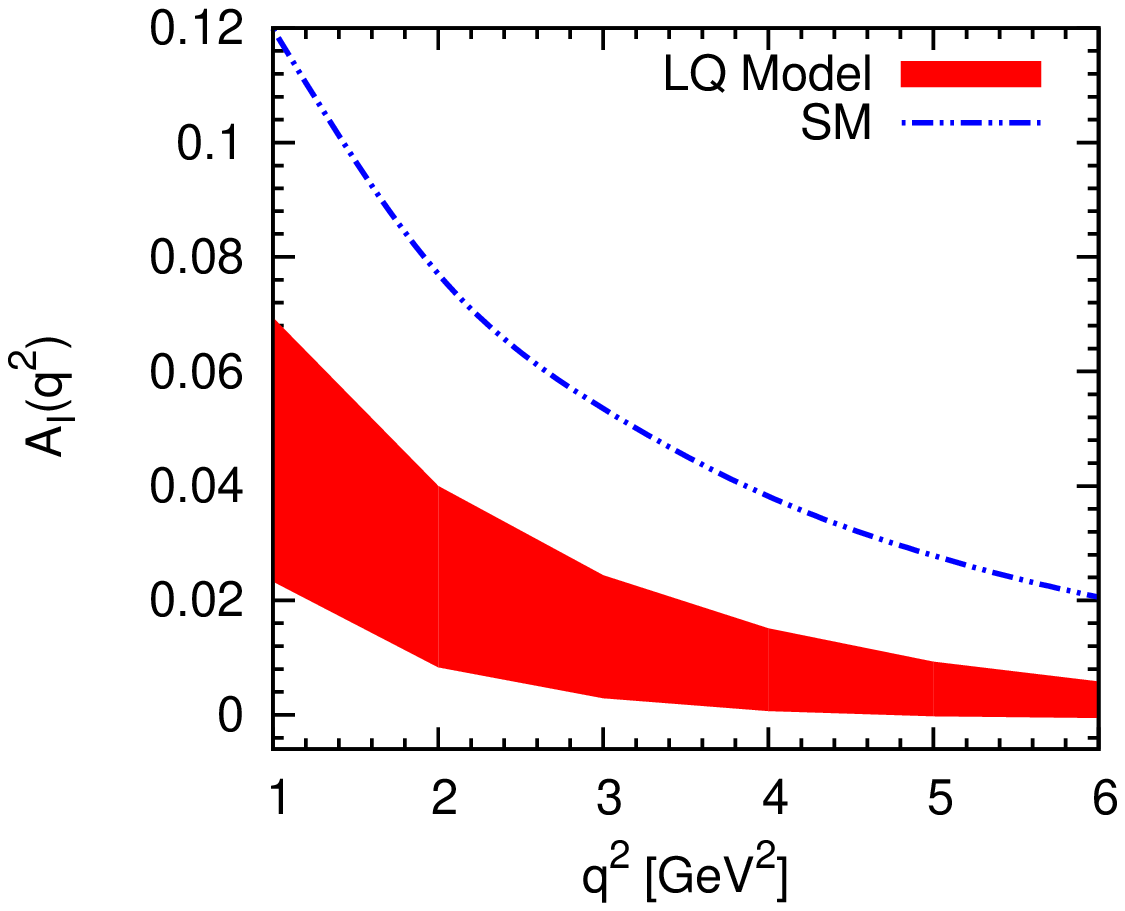}
\hspace{0.2 cm}
\includegraphics[width=8.0 cm,height=6.5cm, clip]{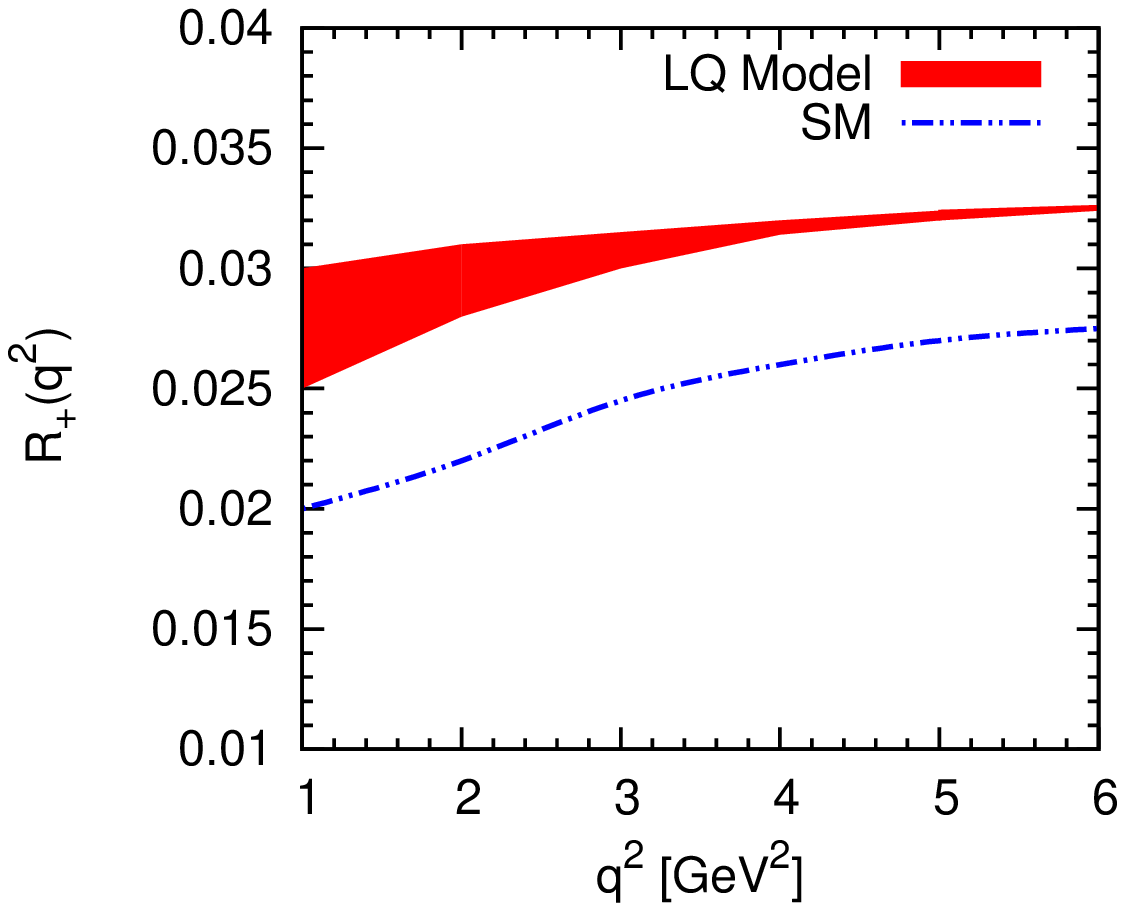}
\caption{The variation of isospin asymmetry for  $B \to \pi \mu^+ \mu^-$ process and $R_+(q^2)$  both in SM and LQ model.}
\end{figure}

\section{Lepton Flavour Violating decays $B_{s,d} \to l_i^+ l_j^-$}

It is very well known that in the standard model the
family lepton numbers ($L_e,~L_\mu, L_\tau)$ are exactly conserved. However, the
experimental observation of neutrino oscillation implies that the family lepton numbers
are no longer conserved quantum numbers and  must be violated.
Due to the  violation of these lepton numbers, flavour changing neutral
current (FCNC) processes in the lepton sector could in principle occur, analogous to the
quark sector. Some examples of FCNC transitions in the lepton sector are: $l_i \to l_j \gamma$,
$l_i \to l_j l_k\bar{l}_k$, $B \to l_i\bar{l}_j$ etc., where $l_i$
is any charged lepton. Although there is
no direct conclusive experimental evidence for such processes that have been observed so far,
but there exist severe constraints on some of these lepton flavour violation (LFV) decay modes \cite{pdg}.
The LFV decays are well studied  in the literature in various beyond the stanadard
model scenarios. Here we would like to investigate the effect of scalar leptoquarks in predicting the
branching ratios for the LFV decays  $B_{s,d} \to l_i^+ {l}_j^-$. These decay modes are previously
investigated in \cite{rm2}.

The effective Hamiltonian for $B_{s,d} \to l_i^+ l_j^-$ process will have similar structure analogous to
$B_{s,d} \to l^+ l^-$, which is given  in the leptoquark model  as
\bea
{\cal H}_{LQ}=  \Big[G_V \Big ( \bar s \gamma^\mu P_L b\Big)~{\bar l}_i
\gamma_\mu l_j +  G_A \Big(\bar s \gamma^\mu P_L b\Big)~{\bar l}_i
\gamma_\mu \gamma_5 l_j \Big]\;,
\eea
where the constants $G_V$ and $G_A$ are given as
\be
G_V =G_A = \frac{\lambda^{j3} {\lambda^{i2}}^* + \lambda^{j2} {\lambda^{i3}}^*}{8M_Y^2}\;.
\ee
Here we have considered the exchange of the  leptoquark as $X(3,2,7/6)$ and for $X(3,2,1/6)$, one will have the chirality-flipped
operators.

This gives the branching ratio as
\bea
{\rm BR}(B_{s,d} \to l_i^+ l_j) &=& \frac{|{\bf p}|}{4 \pi m_B^2}  |F_V f_{B_q}|^2 \biggr[
 (m_j-m_i)^2 \Big(m_B^2 -(m_i+m_j)^2 \Big) \nn\\
&+& (m_j+m_i)^2\Big (m_B^2 -(m_i-m_j)^2 \Big )\biggr]
\eea
where
\be
|{\bf p}|= \frac{\sqrt{(m_B^2 -m_i^2-m_j^2)^2 -4 m_i^2 m_j^2}}{2 m_B}
\ee
is the center-of-mass momentum of the outgoing leptons in the initial $B_{s,d}$ rest frame.

For numerical estimation we need to know the values of the different couplings involved in the expression for branching ratio.
Asuuming the leptoquarks to have full strength coupling to a lepton and a quark of the same generation and its  coupling
with the quarks and leptons of different generations are assumed to be Cabibbo suppressed. We use the values of these couplings
extracted from the leptonic decays $B_{s,d} \to \mu^+ \mu^-$ as the benchmark values and determine the other required couplings assuming
that the couplings between different generation of quarks and leptons follow the simple scaling law, i.e.
$ \lambda^{ij} = \left (m_i/m_j\right )^{1/4} \lambda^{ii}$ with $j>i$. This assumption follows from the fact that in the quark sector the expansion
parameter of the CKM matrix in the Wolfenstein parameterization can be related to the down type quark masses as $\lambda \sim \sqrt{m_d/m_s}$
where as in the lepton sector one can have the same order for $\lambda$ with the relation $\lambda \sim (m_e/m_\mu)^{1/4}$.
With this simple ansatz, the predicted values of the branching ratios for various LFV decays are listed in Table-V, which
are consistent with present experimental upper limits \cite{pdg}.

\begin{table}[htb]
\begin{center}
\vspace*{0.1 true in}
\begin{tabular}{|c|c|c|c|}
\hline
Decay Process & Couplings involved & Predicted BR &~ Expt. Upper limit \cite{pdg}~  \\
\hline
$B_d \to \mu^\pm e^\mp $ &~ $\frac{\lambda^{31} {\lambda^{12}}^*}{M_S^2}$ ~& ~$ \left ( 9.5 \times 10^{-13} - 6.4 \times 10^{-12}\right )$~ &
$< 2.8 \times 10^{-9}$\\

 &~ $\frac{\lambda^{32} {\lambda^{11}}^*}{M_S^2}$ ~& ~$ \left ( 2.0 \times 10^{-10}-1.3 \times 10^{-9}\right ) $~&\\
\hline

$B_d \to \mu^\pm \tau^\mp $ &~ $\frac{\lambda^{32} {\lambda^{13}}^*}{M_S^2}$ ~& ~$ \left (7.5 \times 10^{-10}-5.1
\times 10^{-9} \right )$~&$ <2.2 \times 10^{-5}$ ~\\

 &~ $\frac{\lambda^{33} {\lambda^{12}}^*}{M_S^2}$ ~& ~$  \left (1.3 \times 10^{-8}- 8.5 \times 10^{-8} \right ) $~&~\\
\hline

$B_d \to e^\pm \tau^\mp $ &~ $\frac{\lambda^{31} {\lambda^{13}}^*}{M_S^2}$ ~& ~$ \left (5.2 \times 10^{-11}-3.5 \times 10^{-10} \right ) $~
&$<2.8 \times 10^{-5}$~\\

 &~ $\frac{\lambda^{33} {\lambda^{11}}^*}{M_S^2}$ ~& ~$ \left ( 1.8 \times 10^{-7}- 1.2 \times 10^{-6} \right )$~&\\
\hline

$B_s \to \mu^\pm e^\mp $ &~ $\frac{\lambda^{32} {\lambda^{21}}^*}{M_S^2}$ ~& ~$ < 1.5 \times 10^{-11} $~&$<1.1 \times 10^{-8}$~\\

 &~ $\frac{\lambda^{31} {\lambda^{22}}^*}{M_S^2}$ ~& ~$ < 3.2 \times 10^{-9} $~&\\
\hline

$B_s \to \mu^\pm \tau^\mp $ &~ $\frac{\lambda^{32} {\lambda^{23}}^*}{M_S^2}$ ~& ~$ < 1.2 \times 10^{-8} $~&$-$\\

 &~ $\frac{\lambda^{33} {\lambda^{22}}^*}{M_S^2}$ ~& ~$ < 2.0 \times 10^{-7} $~&\\
\hline
$B_s \to e^\pm \tau^\mp $ &~ $\frac{\lambda^{31} {\lambda^{23}}^*}{M_S^2}$ ~& ~$ < 8.5 \times 10^{-10} $~&$-$\\

 &~ $\frac{\lambda^{33} {\lambda^{21}}^*}{M_S^2}$ ~& ~$ < 2.9 \times 10^{-6} $~&\\

\hline
\end{tabular}
\end{center}
\caption{The predicted  branching ratios for various lepton flavor violating $B_{s,d}$ decays.}
\end{table}


\section{Conclusion}
In this paper we have studied the effect of the scalar leptoquarks in the rare decays of 
$B \to K l^+ l^-$, $B \to \pi \mu^+ \mu^-$ and the lepton flavour violating decays
$B \to l_i^+ l_j^-$. 
We have considered the simple renormalizable leptoquark models in which 
proton decay is prohibited at the tree level. The
leptoquark parameter space has been constrained
using the recent measurements   on
${\rm BR}(B_{s,d} \to \mu^+ \mu^-)$  and the value of ${\rm BR}(\bar B_d^0 \to X_s e^+ e^-)$.
Using such parameters we obtained the bounds on the product of leptoquark couplings and
 then estimated the branching ratios, isospin asymmetries for $B \to K \mu^+ \mu^-$ process.
We found that the observed anomaly of $R_K$ can be explained in the leptoquark model.
This is because  the couplings of leptoquarks are family dependent and one can have lepton flavour
interaction in this model. For $B \to \pi \mu^+ \mu^-$, we have studied the effect of
leptoquarks on branching ratios, CP asymmetry parameters, isospin asymmetry parameter $A_I$ and
$R_+$ parameter which corresponds to the ratio of the branching ratios of $B^+ \to \pi^+ \mu^+ \mu^-$ to  
$B^+ \to K^+ \mu^+ \mu^-$. For $B \to \pi \mu^+ \mu^-$ decays, these observavbles deviate significantly
from their corresponding SM values. We have also obtained the branching ratios for various lepton flavour
violating decays $B \to l_i^+ l_j^-$. Some of these decay modes, e.g., $B_d \to \mu^\pm \tau^\mp$ are
expected to have branching ratios which are within the reach of LHCb, the observation of which
 would provide the hints for possible existence of leptoquarks.

{\bf Acknowledgments}

We would like to thank Science and Engineering Research Board (SERB),
Government of India for financial support through grant No. SB/S2/HEP-017/2013.

\appendix
\section{Amplitude for $B \to P \gamma^*$ process}
Here we will present the expressions for $B \to P \gamma^* $ amplitudes from Refs. \cite{beneke1, beneke2}.
Including corrections ${\cal O}(\alpha_s)$, the $B \to P \gamma^*$ amplitude   in the heavy quark limit is given by
\bea
{\cal T}_P^{(i)} = \xi_P C_P^{(i)} + \zeta_P \sum_{\pm} \int_0^\infty \frac{d \omega}{\omega}
\Phi_{B,\pm}(\omega)\int_0^1 du \phi_P(u) T_{p, \pm}^{(i)}(u, \omega)\;,
\eea
where
\be
\zeta_P = \frac{\pi^2}{N_C}\frac{f_B f_P}{M_B}\;.
\ee
The  expressions for the coefficient functions $C_P^{(i)}$ and ${\cal T}_{P,\pm}^{(i)} $  are given as
\be
C_P^{(i)} = C_P^{(0,i)}+ \frac{\alpha_s C_F}{4 \pi} C_P^{(1,i)}\;,
\ee
\be
T_{P,\pm}^{(i)}(u,\omega)= T_{P,\pm}^{(0,i)}(u, \omega) + \frac{\alpha_s C_F}{4 \pi} T_{P,\pm}^{(1,i)}(u, \omega)\;,
\ee
and $i=t,u$. The $B \to P \gamma^*$ amplitude can be related to the $B \to V_\parallel \gamma^*$ amplitude as
\be
C_P^{(i)} = - C_\parallel^{(i)}, ~~~~~~~T_{P, \pm}^{(i)}(u,\omega) = -T_{\parallel,\pm}^{(i)}(u,\omega)\;,
\ee
The  expressions for $\bar B \to P \gamma^*$ amplitudes are 
\bea
{\cal T}_P^{(t)} &=& \xi_P \left (C_P^{(0,t)}+ \frac{\alpha_s C_F}{4 \pi} \left[C_P^{(f,t)}+C_P^{(nf,t)} \right ]\right )
+ \zeta_P \lambda_{B,-}^{-1} \int du \phi_P(u) \hat T_{P,-}^{(0,t)}\nn\\
&+& \frac{\alpha_s C_F}{4 \pi} \zeta_P \Big( \lambda_{B,+}^{-1} \int du \phi_P(u) \Big[T_{P,+}^{(f,t)}(u)+T_{P,+}^{(nf,t)}(u) \Big]\nn\\
&+&\lambda_{B,-}^{-1} \int du \phi_P(u) \hat T_{P,-}^{(nf,t)}(u) \Big )\;.
\eea
\bea
{\cal T}_P^{(u)} &=& \xi_P \left (C_P^{(0,u)}+ \frac{\alpha_s C_F}{4 \pi} \left[C_P^{(nf,u)} \right ]\right )
+ \zeta_P \lambda_{B,-}^{-1} \int du \phi_P(u) \hat T_{P,-}^{(0,u)}\nn\\
&+& \frac{\alpha_s C_F}{4 \pi} \zeta_P \Big( \lambda_{B,+}^{-1} \int du \phi_P(u) T_{P,+}^{(nf,u)}(u)
+\lambda_{B,-}^{-1} \int du \phi_P(u) \hat T_{P,-}^{(nf,u)}(u) \Big )\;.
\eea
where use has been made
\bea
T_{P,-}^{(0,i)}(u,\omega) = \frac{M_B \omega}{M_B \omega -q^2-i\epsilon}\hat T_{P,-}^{(0,i)}\;,\nn\\
T_{P,-}^{(nf,i)}(u,\omega) = \frac{M_B \omega}{M_B \omega -q^2-i\epsilon}\hat T_{P,-}^{(nf,i)}(u)\;.
\eea
The  form factor terms including ${\cal O}(\alpha_s^0)$ contributions  are
\bea
C_P^{(0,t)} =C_7^{eff} + \frac{M_B}{2 m_b} Y(q^2),~~~~~C_P^{(0,u)}= \frac{M_B}{2 m_b} Y^{(u)}(q^2)
\eea
where
\be
Y^{(u)}(q^2)=\left (\frac{4}{3}C_1+C_2 \right )[h(s,m_c)-h(s,0)]
\ee
The first order corrections $C_P^{(1,i)}$ are divided into a factorizable and a non-factorizable term and can be written as
\be
C_P^{(1,i)}=C_P^{(f,i)}+C_P^{(nf,i)}\;.
\ee
The  factorizable and nonfactorizable terms including ${\cal O}(\alpha_s)$ correcions are 
\bea
C_P^{(f,t)}= \left (\ln \frac{m_b^2}{\mu^2} + 2L + \Delta M \right )C_7^{eff}\;,
\eea
where $L$ and $\Delta M$ are defined in \cite{beneke1,beneke2}.
\bea
C_F C_P^{(nf,t)}= -\bar{C}_2 F_2^{(7)}-C_8^{eff}F_8^{(7)}
-\frac{M_B}{2 m_b}\left[ \bar{C}_2
F_2^{(9)} +2 \bar{C}_1 \left (F_1^{(9)}+ \frac{1}{6}F_2^{(9)} \right )
+C_8^{eff} F_8^{(9)}\right]
\eea
\bea
C_F C_P^{(nf,u)} &=& -\bar{C}_2 \left (F_2^{(7)}+F_{2,u}^{(7)}\right )
-\frac{M_B}{2 m_b}\Big[ \bar{C}_2
\left (F_2^{(9)} +F_{2,u}^{(9)}\right )\nn\\
&+& 2 \bar{C}_1 \left ((F_1^{(9)}+F_{1,u}^{(9)})+ \frac{1}{6}(F_2^{(9)}+F_{(2,u)}^{(9)}) \right )
\Big]
\eea
The longitudinal amplitude receives a  contribution from  weak annihilation topology, where the photon
couples to the spectator quark in the $B$ meson and 
the ${\cal O}(\alpha_s^0)$ contributions to hard spectator scattering from the weak annihilation diagrams are
\bea
\hat T_{P,-}^{(0,t)}=e_q \frac{4 M_B}{m_b} C_q^{34}\;,~~~~~~~~~~~~~~\hat T_{P,-}^{(0,u)} = -e_q \frac{4 M_B}{m_b} C_q^{12}\;,
\eea
where $e_q$ is the charge of the spectator quark   $q=u,d$ in the $B$ meson  and
\bea
C_q^{34}=C_3 +\frac{4}{3}\left (C_4 + 12 C_5+16 C_6\right ),~~~~~~~~C_q^{12}=3 \delta_{qu} C_2 -\delta_{qd}\left (\frac{4}{3} C_1+C_2 \right )\;.
\eea
The first order corrections can also be divided into a factorizable and a nonfactorizable term as
\bea
{\cal T}_{P,\pm}^{(1,i)}={\cal T}_{P,\pm}^{(f,i)} \pm {\cal T}_{P,\pm}^{(nf,i)}\;.
\eea
The  factorizable and nonfactorizable terms including ${\cal O}(\alpha_s)$ corrections to the hard spectator scattering term are given by
\bea
T_{P,+}^{(f,t)}(u) = -C_7^{eff} \frac{4 M_B}{\bar u E}
\eea
\bea
T_{P,+}^{(nf,t)}(u) &=& - \frac{M_B}{m_b} \Big[e_u t_\parallel(u.m_c)(\bar{C}_2 +\bar{C}_4 - \bar{C}_6  ) \nn\\
&+& e_d t_\parallel (u, m_b) (\bar{C}_3 +\bar{C}_4 -\bar{C}_6) +e_d t_\parallel (u,0) \bar{C}_3 \Big]
\eea
\bea
T_{P,+}^{(nf,u)}(u)=-e_u\frac{M_B}{m_b}\left (C_2 -\frac{1}{6}C_1 \right )[t_\parallel(u,m_c) - t_\parallel(u,0)]\;.
\eea

\bea
\hat{T}_{P,-}^{(nf,t)}(u) &=& -e_q \Big[\frac{8C_8^{eff}}{\bar u + u q^2/M_B^2}+\frac{6M_B}{m_b}
\Big\{h(\bar u M_B^2 +uq^2, m_c)(\bar{C}_2 +\bar{C}_4 + \bar{C}_6 )\nn\\
&+& h(\bar u M_B^2 +uq^2, m_b)(\bar{C}_3 +\bar{C}_4 + \bar{C}_6 )
+h(\bar u M_B^2 +uq^2, 0)(\bar{C}_3 +3\bar{C}_4 + 3\bar{C}_6 )\nn\\
&-& \frac{8}{27}(\bar{C}_3 - \bar{C}_5 -15 \bar{C}_6 )\Big\}     \Big]
\eea

\bea
\hat{T}_{P,-}^{(nf,t)}(u) &=& -e_q \frac{6M_B}{m_b}\left (C_2 - \frac{1}{6} C_1 \right )
\Big[h(\bar u M_B^2 +uq^2, m_c) -h(\bar u M_B^2 +uq^2, 0)\Big]
\eea
where ${\bar C}_i$ (for $i=1,\cdots, 6)$ are defined by
\bea
\bar{C}_1 & = & \frac{1}{2} C_1 \;, \nn\\
\bar{C}_2 & = &C_2- \frac{1}{6} C_1 \;, \nn\\
\bar{C}_3 & = & C_3 -\frac{1}{6} C_4 + 16 C_5 - \frac{8}{3} C_6 \;, \nn\\
\bar{C}_4 & = & \frac{1}{2} C_4+ 8C_6 \;, \nn\\
\bar{C}_5 & = & C_3 -\frac{1}{6} C_4 +4 C_5 - \frac{2}{3} C_6\;, \nn\\
\bar{C}_6 & = & \frac{1}{2} C_4 + 2 C_6\;.
\eea

\end{document}